\documentclass{aa}
\usepackage{txfonts}
\usepackage[utf8]{inputenc}
\usepackage[german,english]{babel}
\usepackage{natbib}
\usepackage{graphicx}
\usepackage{longtable}
\usepackage{lmodern}
\usepackage{amstext}
\usepackage{fp}
\usepackage{url}

\newcommand{\Pm}{\langle P_1 \rangle}
\newcommand{\oldsample}{24,124 } % old sample
\newcommand{\ngood}{18,691 } % good stars (green dots) as defined in Sect. 3.1.1.
\newcommand{\npuls}{454 } % super stable stars (red dots) as defined in Sect. 3.1.1.
\newcommand{\noth}{625 } % very stable stars (blue dots) as defined in Sect. 3.1.1.
\newcommand{\nstable}{1079 } % total number of stable stars, thereof 545 2nd period stars
\newcommand{\nPQ}{17,674 } % quarters matching McQuillan (2014)
\newcommand{\nPQperc}{98.9} % percentage
\newcommand{\nPQalias}{0.5} % percentage

\newcommand{\nPfull}{15,082 } % full matching McQuillan (2014)
\newcommand{\nPfullperc}{97.4} % percentage
\newcommand{\nPfullalias}{1.4} % percentage

\newcommand{\nPseg}{16,443 } % segments matching McQuillan (2014)
\newcommand{\nPsegperc}{96.0} % percentage
\newcommand{\nPsegalias}{2.3} % percentage
\newcommand{\nPcomp}{18,599 } % combined periods matching McQuillan (2014)
\newcommand{\nPcompperc}{97.6} % = 18148/18599
\newcommand{\nPcompround}{18,500 } % rounded number used in the abstract
\newcommand{\nDRround}{12,300 } % rounded number used in the abstract
\newcommand{\ngoodage}{17,623 } % B07 ages between 100--10,000 Myr (97.2%)
\newcommand{\ngoodageMattperc}{90.7} % B07 ages between 100--4000 Myr
\newcommand{\badoldperc}{0.62}
\newcommand{\badyoungperc}{2.2}

\newcommand{\youngerthanSun}{93.0} % B07 ages between 100--4500 Myr
\newcommand{\ngoodageround}{17,000 } % ages between 100 Myr and 10 Gyr using any relation
\begin{document}
%%%%%%%%%%%%%%%%%% HEADER %%%%%%%%%%%%%%%%%%%%%%%%
\title{Rotation, differential rotation, and gyrochronology of active Kepler stars}

\author{Timo Reinhold\inst{1}, Laurent Gizon\inst{1,2}}

\offprints{T. Reinhold, \\ \email{reinhold@astro.physik.uni-goettingen.de} }

\institute{Institut f\"ur Astrophysik, Georg-August-Universit\"at G\"ottingen, 
  37077 G\"ottingen, Germany \and Max-Planck-Institut f\"ur Sonnensystemforschung,  
  Justus-von-Liebig-Weg 3, 37077 G\"ottingen, Germany}

\date{Received day month year / Accepted day month year}

\abstract
{% Context
Apart from the discovery of hundreds of exoplanets, the high-precision photometry from the
CoRoT and Kepler satellites has led to measurements of surface rotation periods for tens
of thousands of stars, which can potentially be used to infer stellar ages via 
gyrochronology.
}
{% Aims
Our main goal is to derive ages of thousands of field stars using consistent rotation
period measurements derived by different methods. Multiple rotation periods are
interpreted as surface differential rotation (DR). We study the dependence of DR with
rotation period and effective temperature.
}
{% Methods
We re-analyze the sample of \oldsample Kepler stars from Reinhold~et~al.~(2013) using
different approaches based on the Lomb-Scargle periodogram. Each quarter (Q1--Q14) is
treated individually using a prewhitening approach. Additionally, the full time series,
and different segments thereof are analyzed.
}
{% Results
For more than \nPcompround stars our results are consistent with the rotation periods from
McQuillan~et~al.~(2014). Thereof, more than \nDRround stars show multiple significant
peaks, which we interpret as DR. Dependencies of the DR with rotation period and effective
temperature, as shown in Reinhold~et~al.~(2013), could be confirmed, e.g. the relative DR
increases with rotation period. Gyrochronology ages between 100\,Myr and 10\,Gyr were 
derived for more than \ngoodageround stars using different gyrochronology relations, most 
of them with uncertainties dominated by period variations. We find a bimodal age 
distribution for $T_{\rm eff}$ between 3200--4700\,K. The derived ages reveal an empirical
activity-age relation using photometric variability as stellar activity proxy. 
Additionally, we found \nstable stars with extremely stable (mostly short) periods. Half
of these periods may be associated with rotation stabilized by non-eclipsing companions, 
the other half might be due to pulsations.
}
{% Conclusions
The derived gyrochronology ages are well constrained since more than 
$\sim$\,\youngerthanSun\,\% of the stars seem to be younger than the Sun where 
calibration is most reliable. Explaining the bimodality in the age distribution is 
challenging, and limits accurate stellar age predictions. The relation between activity 
and age is interesting, and requires further investigation. The existence of cool stars 
with almost constant rotation period over more than three years of observation might be 
explained by synchronization with stellar companions, or a dynamo mechanism keeping the 
spot configurations extremely stable.
}

\titlerunning{Rotation, differential rotation, and gyrochronology of active Kepler stars}
\authorrunning{Timo Reinhold, Laurent Gizon}
\maketitle

\section{Introduction}\label{intro}
% rotation periods
Encouraged by high-precision photometry of the CoRoT and Kepler satellites, measurements 
of stellar rotation periods have become numerous in recent years, unprecedented in their
number and accuracy \citep{Meibom2011_Kepler, Affer2012, McQuillan_Mdwarfs, Nielsen2013,
McQuillan_KOI, Walkowicz2013, Reinhold2013, McQuillan2014}. 
% magnetic braking 
Owing to magnetic braking, the rotation period of a star is linked to its age. Stellar
winds carry away charged particles along the magnetic field lines. As time proceeds, more
and more material is dissipated, making the star slower and slower. This angular momentum
loss over the stars' lifetime results in the fact that young stars rotate faster than old
ones, on average.

% Skumanich: activity-rotation-age relation
A relation between stellar age and rotation period was first shown by 
\citet{Skumanich1972}. This author further demonstrated that the level of chromospheric
activity depends on the stellar age. Succeeding measurements at the Mount Wilson
Observatory corroborated the relations between chromospheric activity index $R'_{\rm HK}$,
the rotation period, and the stellar age \citep{Noyes1984, Soderblom1991}. Additionally,
since the 1980s fast rotators are known to exhibit enhanced X-ray activity
\citep{Pallavicin1981,Pizzolato2003}. These coherencies are now established as
activity-rotation-age relation.

% ages: methods
The stellar age cannot be measured directly but has to be inferred from other quantities.
In principle, activity-age relations can be used to infer stellar ages, although their
accuracy suffers from secular changes of the activity level. 
% isochrone fitting 
A classical method to infer stellar ages is isochrone modeling. The ages of stellar
clusters can be inferred from this method (see, e.g., \citealt{Perryman1998}), provided
some knowledge of effective temperature, luminosity, and metallicity. This method can also
be applied to field stars, however with typical uncertainties on the order of 20--30\,\%.

% Gyrochronology
Over the past years, \textit{gyrochronology} has become a promising method to derive
stellar ages from their rotation periods \citep{Barnes2003, Barnes2007, Mamajek2008,
Meibom2009, CollierCameron2009, James2010, Barnes2010, Delorme2011, Meibom2011}.
\citet{Barnes2003,Barnes2007} collected rotation periods from open clusters of different
ages, and identified two sequences in the color-period diagram, namely the $I$- and
$C$-sequence. Using known cluster ages, this author derived empirical fits to the color
(mass) dependence of the data on the $I$-sequence, according to
\begin{equation}
  P_I(B-V, t) = f(B-V) \, g(t).
\end{equation}
The function $g(t) \propto t^{1/2}$ is the rotation-age relation from
\citet{Skumanich1972}, and $f(B-V)$ an empirical fit to the $(B-V)_0$ color dependence of
the rotation period $P_I$ of the $I$-sequence stars reflecting their braking efficiency.
$(B-V)_0$ colors are used as a substitute for stellar mass instead. Unfortunately,
gyrochronology is poorly calibrated for old stars. \citet{Meibom2011_Kepler} measured
rotation periods for the 1\,Gyr old cluster NGC\,6811 in the Kepler field. The oldest star
used for calibration was the Sun with an age of 4.55\,Gyr. Bridging this large gap
\citet{Meibom2015} recently measured rotation periods of 30 stars in the 2.5\,Gyr old
cluster NGC\,6819, deriving a well-defined period-age-mass relation. At later ages wide
binaries might help constraining the age-rotation relation \citep{Chaname2012}.

% asteroseismology
Another method to determine stellar ages is provided by asteroseismology (see, e.g.,
\citealt{Chaplin2014}). Recently, efforts are underway to calibrate asteroseismology
ages by comparing them to gyrochronology measurements \citep{Garcia2014, Nascimento2014,
Lebreton2014, Angus2015}. % add Nielsen2015 when available
% magnetochronology
Furthermore, \citet{Vidotto2014} found a correlation between the average large-scale
magnetic field strength and the stellar age according to $\langle|B_V|\rangle\propto
t^{-0.655}$ similar to Skumanich's law, calling this method \textit{magnetochronology}.
A review article on various age dating methods was provided by \citet{Soderblom2010}.

% method pros & cons
Gyrochronology relies on age calibration using open clusters of different ages, assuming
their stars to be coeval. Calibration becomes difficult for clusters older than
$\sim$\,1\,Gyr, though. One reason is that rotation periods are difficult to measure for
slowly rotating stars. Furthermore, clusters become disrupted at that age (the degree
strongly depends on the initial stellar cluster density), rendering the cluster membership
of individual stars uncertain. Additionally, gyrochronology relations are calibrated only
for main sequence dwarfs. Bulk rotation period measurements might be largely contaminated
by subgiants \citep{vanSaders2013}, which have started evolving off the main sequence.
Thus, applying gyrochronology relations using subgiant rotation periods might lead to a
mis-classification of their evolutionary state \citep{Dogan2013}.

% DR
Further uncertainties result from the fact that a star is by no means a rigid rotator with
a well-defined rotation period. High-precision instruments like the CoRoT and Kepler
telescopes provide the opportunity to observe multiple rotation periods associated with
latitudinal differential rotation, which was observed for many active stars
\citep{Reinhold2013}. Moreover, spot rotation periods heavily rely on their evolutionary
timescales, which becomes more important for less active stars. Both effects may
contribute large uncertainties to the rotation period used in the color-period fits.

Despite these drawbacks, gyrochronology relations supply a straightforward way to infer
stellar ages of field stars. Other approaches to measure stellar ages are usually
accompanied with large errors. Isochrone methods collapse for binary stars, which are
assumed to be coeval, but do not appear on the same isochrone if the companions exhibit
different masses. Asteroseismology ages strongly depend on some (usually unknown) model
parameters (e.g., metallicity), which can lead to large errors.

% goal of the paper
The knowledge of stellar ages is of utmost interest for galactic formation. We aim to
provide stellar ages for thousands of stars, inferred from mean rotation period
measurements via gyrochronology. Furthermore, we show that multiple significant periods
are quite common among active stars, which we assign to surface differential rotation.
These period fluctuations dominate the age uncertainties. As opposed to this, we also
found a sub-sample of stars almost showing no period variations over more than three
years of observation.

% paper structure
The paper is organized as follows: Sect.~\ref{data} presents the Kepler data products. 
The different approaches used to analyze the data are explained in Sect.~\ref{methods}.
Sect.~\ref{results} contains the results, which are further discussed in 
Sect.~\ref{discussion}. We close with a brief summary of our results in
Sect.~\ref{summary}.

\section{Kepler data}\label{data}
The Kepler satellite provides almost continuous observations of the same field for more
than four years (May 2, 2009 - May 11, 2013). The data is delivered in quarters (Q0--Q17),
each of $\sim$\,90\,d length (Q2--Q16), with exceptions for the commissioning phase Q0
($\sim$\,10\,d), and Q1 ($\sim$\,33\,d). Unfortunately, observations stopped after one
month of Q17 due to a failure of the third reaction wheel. This amazing amount of Kepler
data is publicly available and can be downloaded from the MAST
archive\footnote{http://archive.stsci.edu/pub/kepler/lightcurves/tarfiles/}.

Kepler data has been processed by different pipelines so far, starting with the Presearch
Data Conditioning pipeline (PDC), designed to detect planetary signals. This pipeline was
changed to the so-called PDC-MAP (Maximum A Posteriori) pipeline \citep{Stumpe2012,
Smith2012} because the PDC pipeline coarsely removed stellar variability signals.
Recently, all Kepler data has been reprocessed by the PDC-msMAP (multiscale MAP) pipeline
\citep{Stumpe2014}. This new version applies a 20-days high-pass filter intending to
detect smaller planets in the data. Thus, this pipeline version is not suitable to look
for stellar variability with a broad range of rotation periods since it diminishes
stellar signals of slow rotators. Unfortunately, data reduced by previous pipeline
versions are not publicly available anymore.

In \citet{Reinhold2013} we analyzed $\sim$\,40,000 active stars with a variability range
$R_{\rm var} > 0.3\,\%$ (for definition see \citealt{Basri2010,Basri2011}), only using Q3
data. In $\sim$\,24,000 stars a clear rotation period was detected, and in $\sim$\,18,000
stars a second period was detected, which was assigned to surface differential rotation.
Starting from this sample, we extend our analysis to all available data. Our goal is to
detect consistent rotation periods throughout the quarters, and to refine previous
differential rotation measurements, exploiting the much higher frequency resolution
thanks to the longer time span.

We do not use the much larger sample of 34,030 stars with measured rotation periods from
\citet{McQuillan2014} because we are mostly interested in measuring DR. Although 20,009
stars of our sample are contained in the sample of \citet{McQuillan2014}, the remaining
stars either do not belong to the periodic sample of \citet{McQuillan2014}, or have a
smaller average variability range ($R_{\rm var} < 0.3\,\%$), rendering them unsuitable for
DR measurements.

In this work we only use Q1--Q14 PDC-MAP data (version 2.1 for Q1--Q4, Q9--Q11 and
version 3.0 for Q5--Q8, Q12--Q14). Q0 data was discarded because of its short time span
and the much lower number of monitored targets. From Q15 on, only PDC-msMAP data was
available, which is unsuitable for our purposes as explained above.

Since we are only interested in rotation-induced stellar variability, we have to exclude
targets showing other kinds of periodic variability. To reduce the number of false
positives, i.e., periodic variability not related to stellar rotation, we discarded 17
eclipsing binaries\footnote{http://keplerebs.villanova.edu/}, 878 planetary
candidates\footnote{http://archive.stsci.edu/kepler/koi/search.php}, 2 RR Lyrae stars
\citep{Kolenberg2010, Szabo2010, Benko2010, Guggenberger2012, Nemec2011, Moskalik2012,
Molnar2012, Nemec2013}, and 84 $\gamma$~Doradus and $\delta$~Scuti stars
\citep{Tkachenko2013,Ulusoy2014, Balona2011, Uytterhoeven2011, Lampens2013, Balona2014}.
In total, 981 stars were discarded, leaving 23,143 targets, which are analyzed as
described in the following section.

\section{Methods}\label{methods}
After excluding binarity stars and pulsators, we are interested in the stability of the
rotational modulation. Many effects can change the shape of a light curve dominated by
star spots rotating in and out of view, e.g., spots are created, others disappear while
rotation takes place. The number of spots and their sizes are usually unknown.
Differential rotation further hampers the detection of stable rotation periods. Sun spots
change their preferred latitude of occurrence during the solar activity cycle, and
therefore their rotation rate. But cyclic variability is also expected in other active
stars. Moreover, Kepler suffers from instrumental effects, which are corrected by the
pipeline automatically. Improper correction can mimic long-term (periodic) variability.
The Kepler satellite rolls between consecutive quarters to re-orientate its solar arrays.
Hence, in each quarter stars fall on different CCDs with different sensitivity.

To account for all these effects, we analyze the data of the same star in different ways.
First, we apply our analysis to each quarter individually. By comparing periods of
individual quarters among themselves, we shrink our initial sample to find stable rotation
periods among many quarters. After that, we stitch together data from all quarters for
each star in our sample, and apply a slightly different period search to the full light
curve. In a last step, we analyze different segments of the full light curve, because some
effects, (e.g., differential rotation) are more visible in certain segments rather than in
the full time series. By comparing the periods returned by each method, we hope to detect
stable rotation periods, which are unlikely caused by instrumental systematics.

\subsection{Analysis of individual quarters}\label{quarters}
To detect periodic signals in the data we use the Lomb-Scargle periodogram in a
prewhitening approach as described in \citet{Reinhold2013}. Each quarter is analyzed
individually and in the same way. The analysis method in briefly summarized below. For
details we refer the reader to the paper mentioned above.

For each star and each quarter we compute the variability range $R_{\rm var,Q}$. If
$R_{\rm var,Q} > 0.3\,\%$ holds for a certain quarter, we compute five Lomb-Scargle
periodograms in a successive prewhitening approach. Since computing the Lomb-Scargle
periodogram is equivalent to fitting a sine wave to the data, each prewhitening step
yields a set of sine fit parameters (period $P_k$, amplitude $a_k$, phase $\phi_k$, and a
total offset c) for $k=1,...,5$. The parameters which belong to the highest peak in the
periodogram yield the best parameters to fit the data. We use the returned values as
initial parameters for a global sine fit to the data $(t,y)$ according to
\begin{equation}\label{fit}
 y_{fit} = \sum_{k=1}^5 a_k \sin(\frac{2\pi}{P_k}\,t -\phi_k) +c,
\end{equation}
This global sine wave is fit to the data through $\chi^2$-minimization. To save
computation time each light curve was binned to two hours cadences. The period associated
to the highest peak in the Lomb-Scargle periodogram is called $P_1$, and represents the
most significant period in the data. 

% alias detection
In some cases spots are located on opposite sides of the star, and only half of the true
rotation period is detected. To reduce this number of \textit{alias} periods we compare
$P_1$ to the set of periods $P_k$ and their corresponding peak heights $h_k:=h(P_k)$. If
it holds
\begin{equation}\label{alias}
  |P_1 - P_k/2| \leq 0.05\,P_1 \quad \text{and} \quad h_k > 0.5\,h_1,
\end{equation}
then the period $P_k$ is likely the true rotation period. In case $P_1$ already was the
correct period, we check if there are periods $P_k$ satisfying
\begin{equation}\label{alias2}
  |P_1 - 2 P_k| \leq 0.05\,P_1 \quad \text{and} \quad h_k > 0.5\,h_1.
\end{equation}
If that is not the case, the period $P_k$ is used as rotation period, and we call this
rotation period $P_1$.

% look for 2nd period
Since our primary goal is to detect differential rotation, we look for periods adjacent
to $P_1$. A probable second period should satisfy
\begin{equation}\label{P2}
  0.01 \leq |P_1 - P_k|/P_1 \leq 0.30
\end{equation}
The period $P_k$ with the second highest power in the prewhitening process satisfying
Eq.~\ref{P2} is called $P_2$. In the following section we compare period measurements
of individual quarters among themselves.

\subsubsection{Comparison of periods from different quarters}
As discussed at the beginning of Sect.~\ref{methods}, the measured periods $P_1$ may
change from quarter to quarter. That is especially true for stars exhibiting a second
period $P_2$ because spots may have changed their location, size, or occurrence at all.
Besides that, the period $P_2$ might be a spurious detection in some quarters. As a first
step, we look for stars with stable rotation periods throughout the quarters. To "pick and
choose" the best stars shrinks our primary sample, but provides a more reliable measure of
the mean stellar rotation period, and eventually the stars' DR.

\begin{figure*}
  \centering
  \includegraphics[width=17cm]{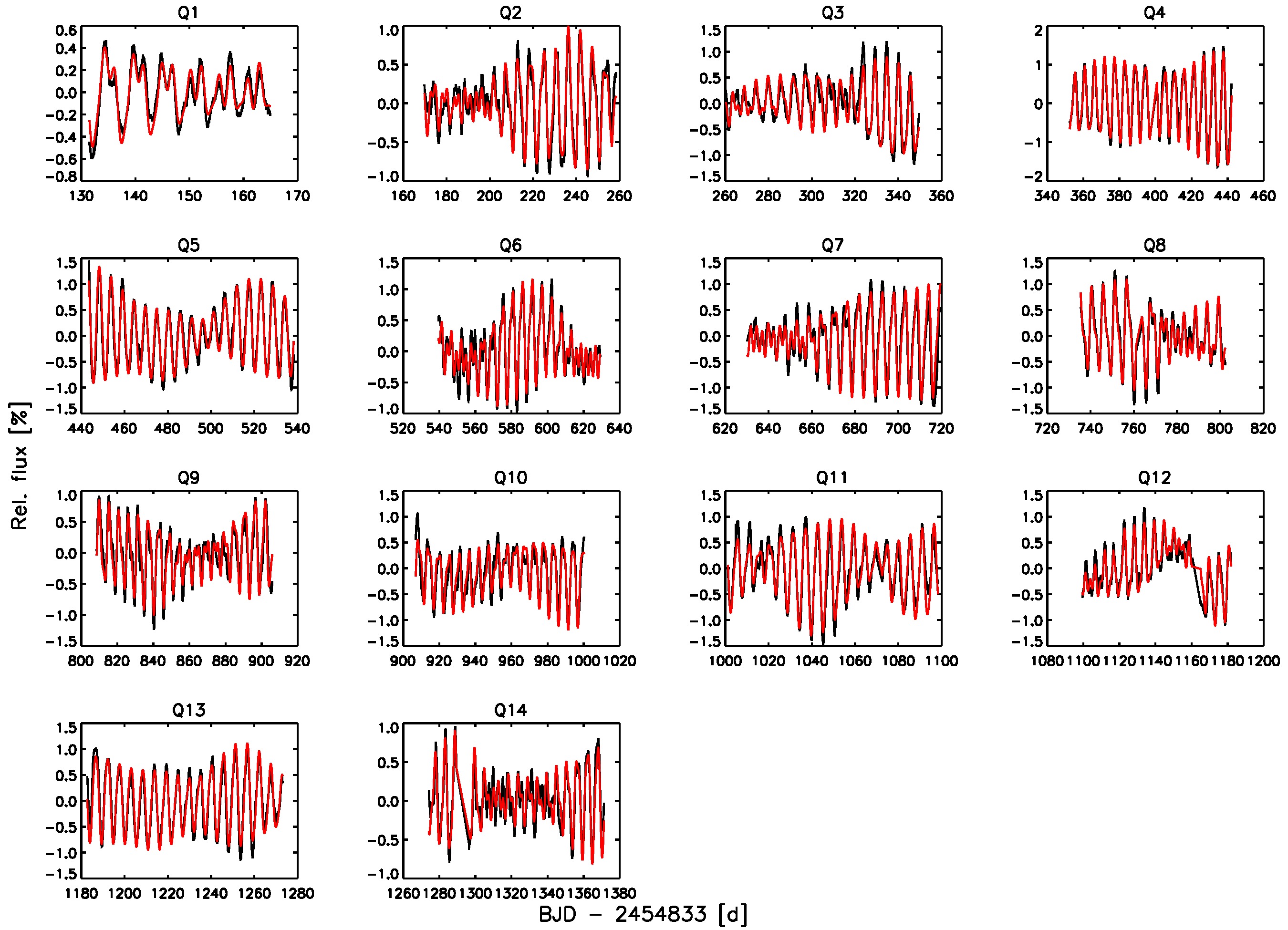}
  \caption{Light curves (black) and sine fits (red) according to Eq.~\ref{fit} of the star
  KIC\,1163579 for all quarters Q1--Q14. Period measurements of this star are given in
  Table~\ref{quarter_table}.}
  \label{lc_quarters}
\end{figure*}

For each star and each quarter $Q=1,...,14$ we compute the variability range\footnote{In
the following, the variability range $R_{\rm var}$ is meant to be the median of the
periods $P_{\rm 1,Q}$ for the quarters Q1--Q14.} $R_{\rm var,Q}$. If $R_{\rm
var,Q}>0.3\,\%$ we apply our prewhitening approach, which yields a period $P_{\rm
1,Q}$. If $R_{\rm var,Q}<0.3\,\%$, or the star was not monitored in a certain quarter, the
period $P_{\rm 1,Q}$ is set to zero. We only consider periods satisfying $\rm 0.5\,d <
P_{1,Q} < 45\,d$. The upper limit accounts for the fact that we want to see at least two
full rotation cycles during the limited quarter length of $\sim$\,90 days. The lower limit
should exclude pulsating stars. We compute the relative deviation of the periods $P_{1,Q}$
from their median according $|P_{\rm 1,Q}-\overline{P_1}|/\overline{P_1}$. Fast rotators
with a median period
$\overline{P_1}<10$\,d are allowed to differ by 10\,\%, stars with longer periods by
20\,\%. From all periods $P_{\rm 1,Q}$ satisfying these criteria, we define a mean
rotation period $\Pm$. If more than 75\,\% of the periods $P_{\rm 1,Q}$ satisfy this
criterion, the star belongs to the so-called "good" sample. The "good" stars are shown in
Fig.~\ref{goodper} compared to previous measurements for the whole sample. We also checked
for alias periods among our set $P_{\rm 1,Q}$ according to Eq.~\ref{alias2}, but without
comparing any peak heights. Periods identified as such were discarded. The star
KIC\,1163579, randomly chosen from our sample, is shown in Fig.~\ref{lc_quarters}, and its
briefing is shown in Table~\ref{quarter_table}. In the following section we concatenate
data from individual quarters to analyze the full light curve.
% Figure example star
% Table example star
\begin{table}
  \centering
  \begin{tabular}{cccc}
\hline\hline
Quarter & $P_1$ [d] & $P_2$ [d] & $R_{\rm var}$ [\%] \\
\hline
Q01 & 5.51 & 4.98 & 0.69 \\
Q02 & 5.71 & 6.06 & 1.41 \\
Q03 & 5.39 & 5.14 & 1.55 \\
Q04 & 5.47 & 6.04 & 2.34 \\
Q05 & 5.34 & 5.74 & 1.78 \\
Q06 & 5.48 & 5.12 & 1.63 \\
Q07 & 5.44 & 5.20 & 1.86 \\
Q08 & 5.30 & 5.81 & 1.78 \\
Q09 & 5.38 & 5.79 & 1.44 \\
Q10 & 5.33 & 5.74 & 1.32 \\
Q11 & 5.37 & 6.02 & 1.79 \\
Q12 & 5.29 & 5.89 & 1.66 \\
Q13 & 5.37 & 5.95 & 1.66 \\
Q14 & 5.64 & 5.26 & 1.17 \\
\hline
\end{tabular}

  \caption{Period measurements of the star KIC\,1163579 from Fig \ref{lc_quarters}. For
  all quarters Q1--Q14 the star was monitored, and the variability range $R_{\rm var,Q} >
  0.3\,\%$. In each quarter a second period was found, and no alias period was detected.
  For the star KIC\,1163579 we find $\Pm=5.43$\,d.}
  \label{quarter_table}
\end{table}

\subsection{Analysis of the full time series}\label{full}
% motivation: why do we need to analyze the full time series?
In the previous section we only considered periods shorter than 45 days, due to the
limited quarter length of $\sim$\,90 days. Stitching together data from all available
quarters is the only way to achieve two major goals: 1) the detection of periods longer
than 45 days, and 2) to obtain a higher frequency resolution in the Lomb-Scargle
periodogram, since the peak width scales with the inverse of the time span.

% long periods...difficult
The first point is difficult to address. As mentioned earlier long-term instrumental
effects are sometimes difficult to distinguish from slow rotation, especially in an
automated period search of a large sample of objects. Furthermore, for this particular
sample rotation periods less than 45 days were measured in \citet{Reinhold2013}.
Measurements of alias periods or spurious detections are possible but expected to be
rare. Nevertheless, we use an upper period limit of 60 days for the analysis of the full
time series.

% higher resolution
The second point is the more interesting one because a higher frequency resolution offers
the detection of individual peaks lying much closer than in the periodograms of the
individual quarters. Thus, the frequency resolution is crucial for the detection of small
values of differential rotation!

% roadmap
To achieve an almost continuous time series of each star we chose the easiest way to
concatenate consecutive quarters by dividing the light curves of each quarter by its
median and subtracting unity. Data outliers were removed as described in
\citet{Garcia2011}. Light curves exhibiting low-frequency trends, i.e., increasing or
decreasing behavior over the full 90 days window, were considered as not properly reduced
by the PDC-MAP pipeline, and therefore discarded.

% only one periodogram
The method we apply to the full time series slightly differs from the previously used sine
fit approach. We only compute one Lomb-Scargle periodogram of the whole time series. Due
to the largely increased amount of data each light curve was binned to six hours cadences
to save computation time. This sampling still enables us to detect short periods down to
half a day according to the Nyquist frequency $f_{\rm Nyq}\approx2\,\text{d}^{-1}$. We
identify the
twenty highest peaks and search for periods within the limits $\rm 0.5\,d \leq P_1 \leq
60\,d$. Furthermore, we force a lower peak height limit of $h_1 > 0.10$ to get some
significance for $P_1$, and check for alias periods according to Eq.~\ref{alias}
and \ref{alias2}.

% no over-sampling
An important point to make is that we do not over-sample the periodogram, in contrast to
previous work \citep{Reinhold2013}. For the analysis of individual quarters oversampling
was necessary because many cases only revealed a single broadened peak, rather than two or
more distinct peaks. Therefore, a fine frequency sampling was necessary to subtract the
correct period in the prewhitening to be able to detect more than one period. Owing to the
much higher frequency resolution here, the periodogram is able to reveal individual peaks.

% outlook
In the following section we try to assign a \textit{significance} to individual peaks. The
goal is to find out which peaks really carry information about different spot rotation
periods, and which are related to spurious detections and/or stochastic effects, e.g.,
emerging and waning active regions.

\subsubsection{Identification of significant peaks}
\begin{figure*}
  \centering
  \includegraphics[width=17cm]{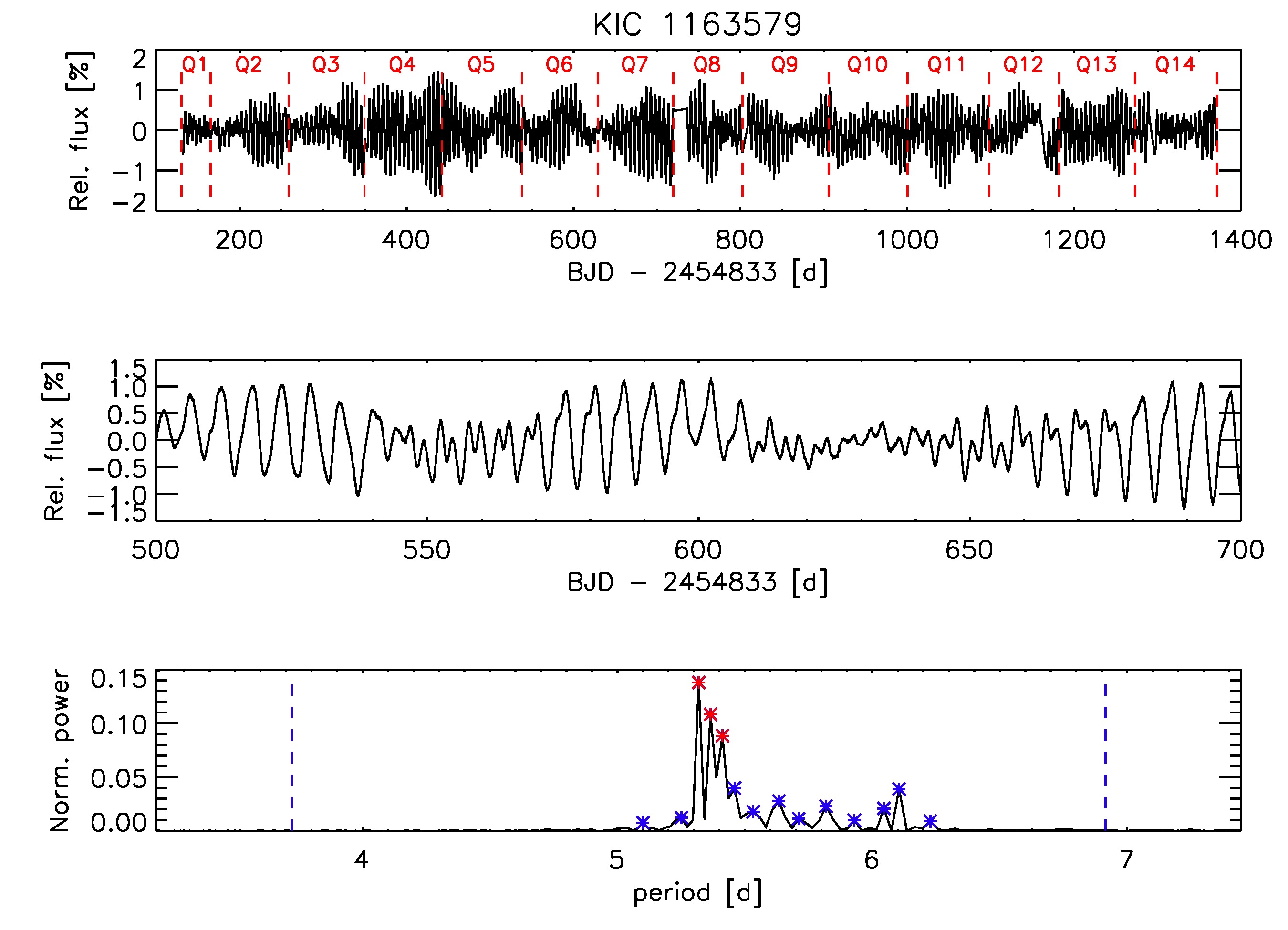}
  \caption{\textit{Upper panel:} Full light curve of the star KIC\,1163579 from quarters
  Q1--Q14. \textit{Middle panel:} Zoom into the above light curve. Periodicity and the
  existence of a second period is clearly visible, revealed by the double-dip structure
  around 550 days. \textit{Lower panel:} Periodogram of the full light curve. Significant
  peaks are marked in red, other peaks in the range of $P_1\pm30\,\%$ (indicated by the
  dashed blue lines) are marked in blue.}
  \label{lc_full}
\end{figure*}

\begin{figure}
  \resizebox{\hsize}{!}{\includegraphics{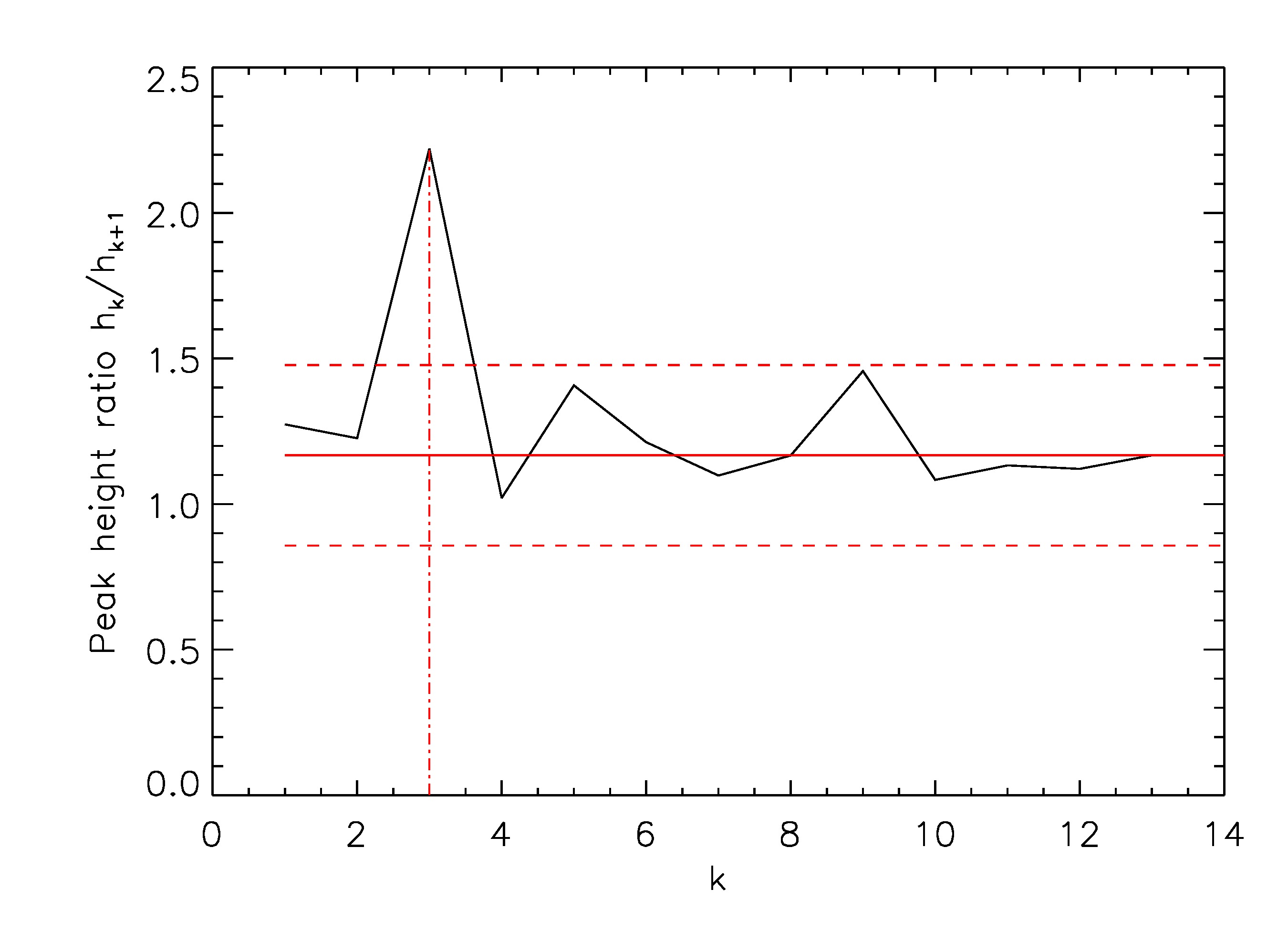}}
  \caption{The peak height ratio (PHR) $h_k/h_{k+1}$ versus peak number $k$ of the
  periodogram from Fig.~\ref{lc_full}. The solid red line shows the median PHR, and the
  dashed red lines mark the $\pm 1\sigma$ levels. At $k=3$ the deviation from the median
  is at maximum, meaning that three significant peaks were found by this method. These
  are marked by red asterisks in Fig.~\ref{lc_full}.}
  \label{PHR}
\end{figure}

Analogous to Eq.~\ref{P2} we search for periods $P_k$ within 30\,\% of $P_1$. We sort the
peak heights in descending order according to $h_1 > h_2 > ...$ and compute the so-called
\textit{peak height ratio} (PHR) $h_k/h_{k+1}$. This method is illustrated based on the
example star KIC\,1163579 in Fig.~\ref{PHR}. The idea is to detect peaks with comparable
heights, and to see where the peak height drops to a significantly lower value from one
peak to the other. Hence, we compute the median of the PHR and search for the maximum
deviation from this value. The related index $k=k_{\rm max}$ yields the number of
significant peaks. In Fig.~\ref{PHR} the median is shown as solid red line, and
the dashed red lines indicate the $\pm1\sigma$ region. For this example, it is evident
that for $k_{\rm max}=3$ the PHR is at maximum, which means that the peak height of $P_3$
is much higher than that of $P_4$, compared to all peak heights within $P_1\pm30\,\%$. All
periods $P_k$ with $k \leq k_{\rm max}$ are considered as significant, the others are
discarded. The lower panel of Fig.~\ref{lc_full} shows the periodogram of the active star
KIC\,1163579 with the significant peaks found by this method marked in red.

\subsection{Analysis of segments of the full time series}\label{segments}
% why do we need this method?
This section contains another approach we used to extract information about different
periods in the data. The primary peak in the periodogram of the full time series
represents the best sine fit period over the full observing time. This period must be
interpreted as an average rotation period, because it fits certain parts of the light
curve better than others. Adjacent peaks with less power are also needed to properly fit
the data, and these periods may be more \textit{visible} in certain segments than in the
full light curve. 

% adjacent peaks: where do they come from?
Minor peaks adjacent to $P_1$ (e.g., peaks marked by blue asterisks in the lower panel
of Fig.~\ref{lc_full}) result from various phenomena: differential rotation, spot
evolution, and data reduction. Differentiating between these effects is almost impossible,
but we assume that DR is the dominating effect in most cases. Spot evolution might be the
strongest contributor of spurious detections of DR, where multiple peaks cannot be
associated to spots rotating at different latitudes. Analyzing different segments thus
becomes important, since the evolution of individual spots does not occur continuously in
time, but may affect certain parts of the full light curve stronger than others. Another
problem might be caused by stitching together individual quarters, where the end of one
quarter and the beginning of the consecutive one do not agree with the overall variability
pattern. This problem is intrinsic to the PDC-MAP pipeline, which was designed to remove
instrumental effects from individual quarters, and not to conserve the overall variability
pattern over the total observing time. 

% segments & method explained
To overcome these effects we analyze segments with different lengths of the full light
curve, aiming to detect periods that are stable over different time scales. In the
following, segments are defined as concatenated quarters, starting with data from Q1--Q2,
and adding the subsequent quarter in the next step, i.e., the second segment contains data
from Q1--Q3, and so on, with the last segment being the full light curve Q1--Q14. We
compute the periodogram of each segment\footnote{For the individual segments we use the
normalization from Eq.~22 in \citet{Zechmeister2009}.}, interpolate it onto the frequency
grid of the periodogram of the full light curve\footnote{This is necessary because
periodograms of different segments possess a different frequency resolution.}, and add up
the logarithmic powers of all segments. This leads to extremely clear periodograms, likely
canceling spurious detections.

\begin{figure*}
  \centering
  \includegraphics[width=17cm]{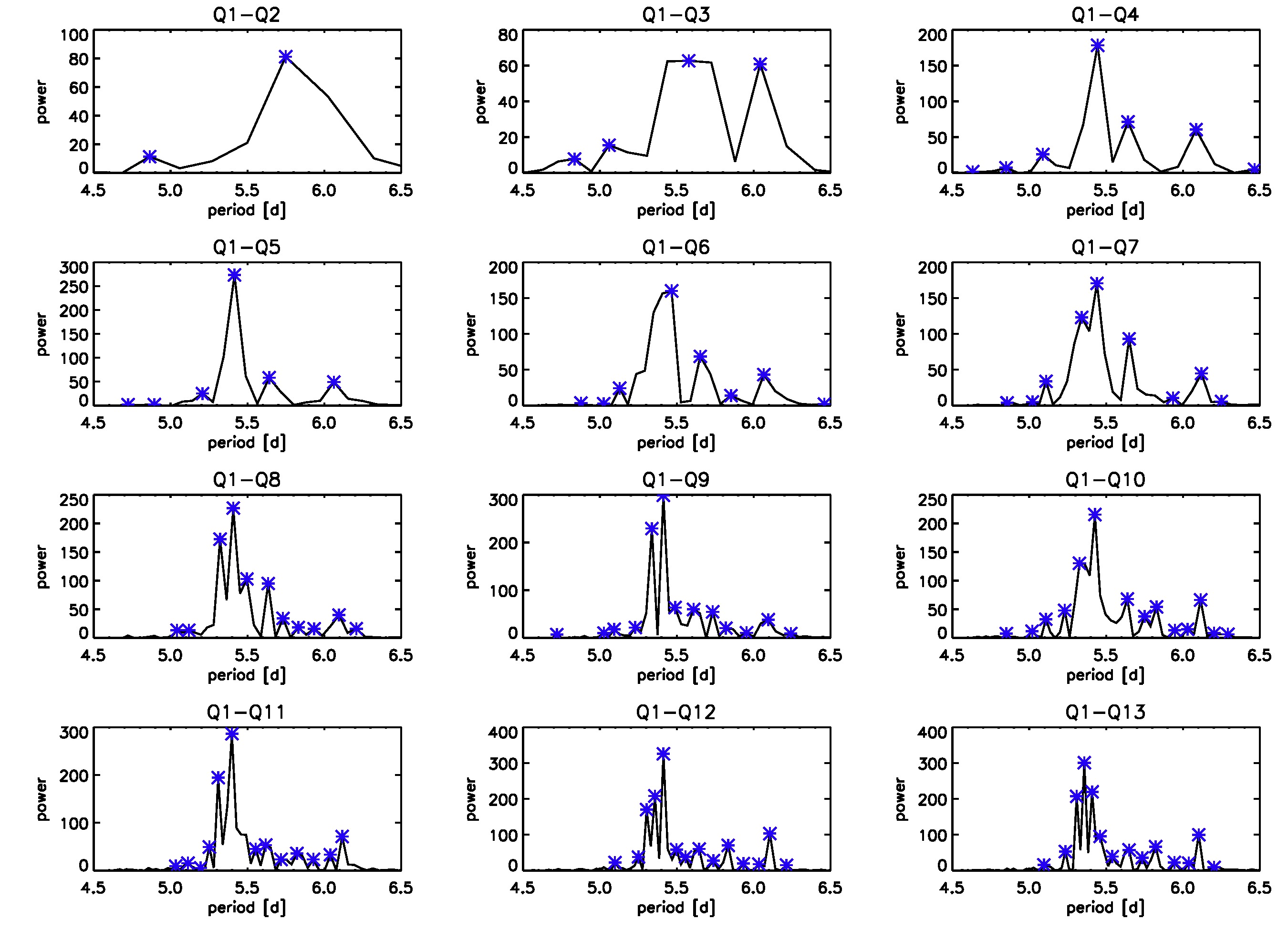}
  \caption{Periodograms of the segments of KIC\,1163579. Periodicity is clearly visible
  between 4.5--6.5~days. The location of the main peak around 5.4~days does not change
  much, whereas the number of minor peaks increases toward longer segments, owing to
  the higher frequency resolution.}
  \label{perdgms_segments}
\end{figure*}

\begin{figure}
  \resizebox{\hsize}{!}{\includegraphics{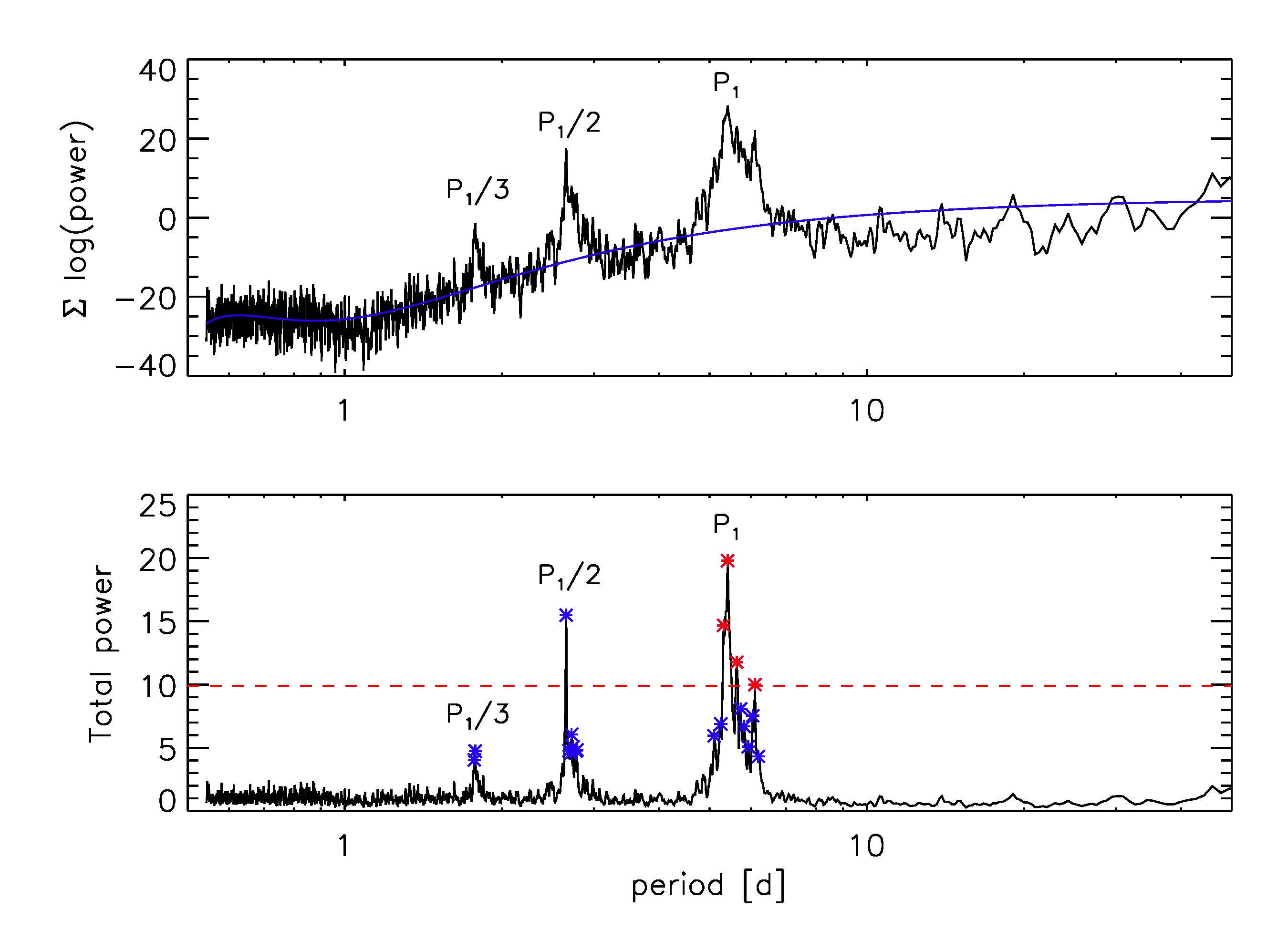}}
  \caption{\textit{Upper panel:} Summed up logarithmic powers of the periodograms of the 
  individual segments from Fig.~\ref{perdgms_segments}. The solid blue curve shows a
  fourth order polynomial fit. \textit{Lower panel:} Exponential power of the upper panel
  with the polynomial fit subtracted. Peaks above the arbitrarily chosen significance
  limit of $0.5\,h_1$ (dashed red line) are considered as significant, and used as a
  measure of DR in the following. In this picture the harmonics ($P_1/2$ and $P_1/3$) are
  also more pronounced.}
  \label{perdgms_tot}
\end{figure}

% example 
Fig.~\ref{perdgms_segments} shows periodograms of the defined segments of the star
KIC\,1163579, clearly revealing multiple periods between 5--6~days in all segments. The
main period crystallizes out around 5.4~days. Each periodogram reveals a period around
6.1~days, which was also found in the periodogram of the full time series
(s. Fig.~\ref{lc_full}), but was not considered as a significant period by our method.
Thus, the measured value of DR might be largely underestimated in this case. To account
for such periods with minor power, we sum up the logarithmic power of all segments, which
is shown in the upper panel of Fig.~\ref{perdgms_tot}. The main power is visible around
$P_1=5.4$~days, and also the first and second harmonic being the half and third of this
period. In certain cases it occurs that the first harmonic, $P_1/2$, has the second
highest or even the highest power in the periodogram. Since we are interested in periods
adjacent to $P_1$ we subtract a fourth order polynomial (solid blue line in
Fig.~\ref{perdgms_tot}), and exponentiate the difference, which is shown in the lower
panel of Fig.~\ref{perdgms_tot}. This procedure leads to extremely clear periodograms.
Periods lying with $\pm30\,\%$ of $P_1$ with powers $>0.5\,h_1$ are considered as
significant. The dashed red line indicates the arbitrarily chosen significance limit. The
period around 6.1~days survives this procedure, and will contribute to the DR measure in
Sect.~\ref{DR}. Results combining our three different approaches are presented in the
following section.

\section{Results}\label{results}
% overview
We present rotation periods of more than \nPcompround stars using the different
approaches from Sects.~\ref{quarters}--\ref{segments}. Based on these periods we search
for differential rotation in Sect.~\ref{DR}. Mean rotation periods together with their
uncertainties are used to infer stellar ages through different gyrochronology relations in
Sect.~\ref{stellar_ages}. Finally, Sect.~\ref{stable} is dedicated to a small fraction of
stars exhibiting rotation periods very stable in time.

\subsection{Rotation periods}\label{periods}
\begin{figure*}
  \centering
  \includegraphics[width=17cm]{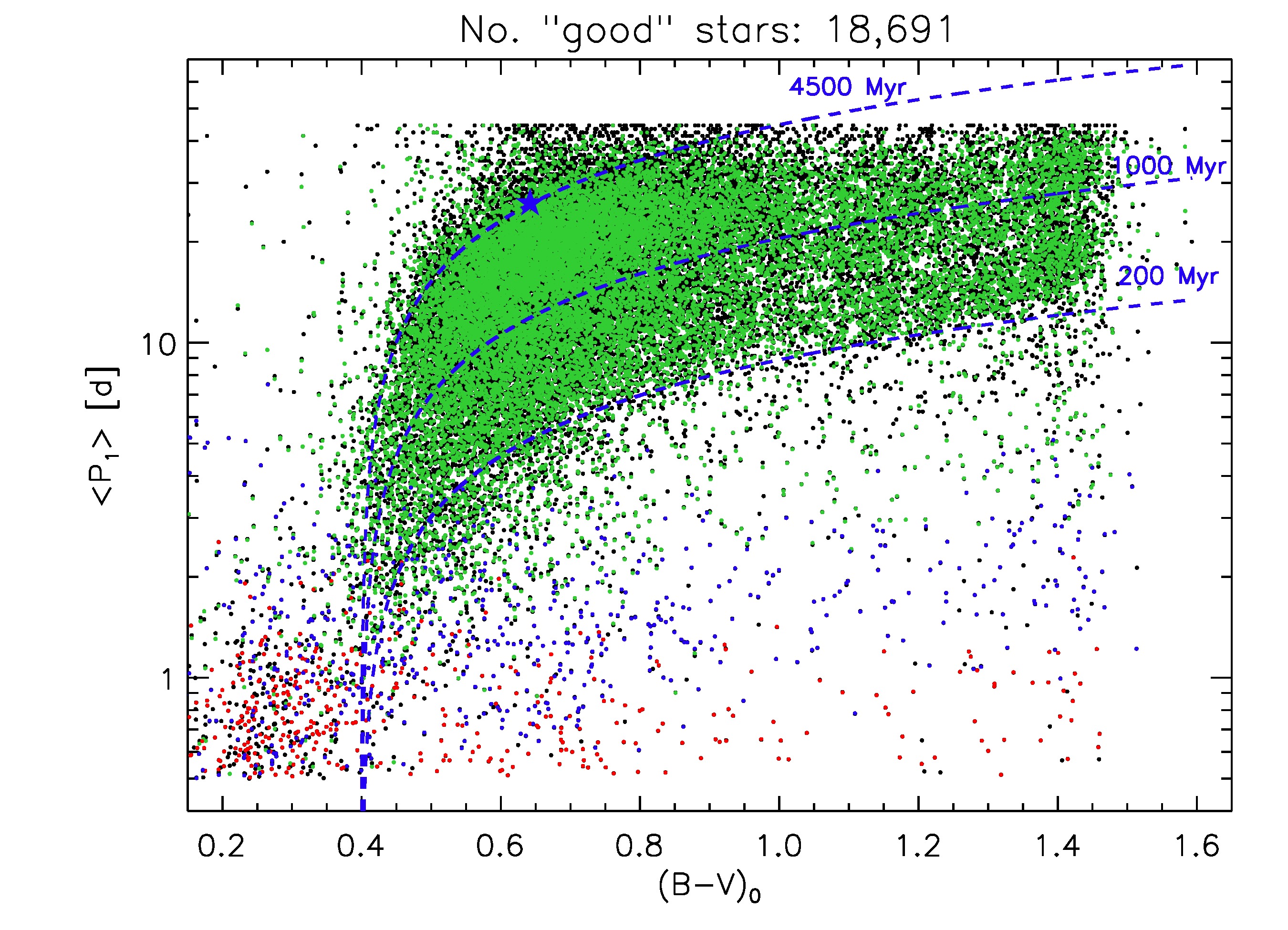}
  \caption{Mean rotation periods $\Pm$ averaged over the quarters Q1--Q14 plotted against
  color $(B-V)_0$. Previous measurements \citep{Reinhold2013} only using Q3 data are shown
  in black, "good" measurements surviving the criteria described in Sect.~\ref{quarters}
  are shown in green. Red and blue dots show stars with periods very stable in time,
  which are discussed separately in Sect.~\ref{stable}. The dashed blue lines show
  isochrones from \citet{Barnes2007}, and the blue star marks the position of the Sun.}
  \label{goodper}
\end{figure*}

% Rotation periods from quarters
In Fig.~\ref{goodper} we present the mean rotation period $\Pm$, averaged over the
quarters Q1--Q14, versus color $(B-V)_0$. The black dots show previous measurements from
\citet{Reinhold2013}, only using Q3 data. Green dots mark the \ngood so-called "good"
stars (s. Sect.~\ref{quarters}), and the \npuls red and \noth blue dots indicate very
stable rotators, which are discussed separately in Sect.~\ref{stable}. The dashed blue
lines show isochrones from \citet{Barnes2007} and the blue star marks the position of the
Sun. Around $(B-V)_0=0.4$ rotational braking due to magnetized winds becomes efficient.
The 4500\,Myr isochrone acts as an upper envelope to our measurements up to
$(B-V)_0\simeq1.0$. Coeval stars redder than $(B-V)_0=1.0$ with supposably longer periods
are missing due to the limited quarter length of $\sim$\,90 days. A lower envelope to the
rotation period distribution is given by the 200\,Myr isochrone. Stars below this curve
are either younger than 200\,Myr, or are not suitable for the use of gyrochronology. The
latter is also true for stars bluer than $(B-V)_0=0.4$ (s. Sect.~\ref{gyro}). A second
group of stars immediately leaps to the eye with short periods between 0.5--2 days and
$(B-V)_0<0.4$. Most of these stars exhibit periods very stable in time, which we discuss
separately in Sect.~\ref{stable}. Comparing the black and green dots it is evident that
the period measurements have been considerably improved by incorporating more data.

\begin{figure*}
  \centering
  \includegraphics[width=17cm]{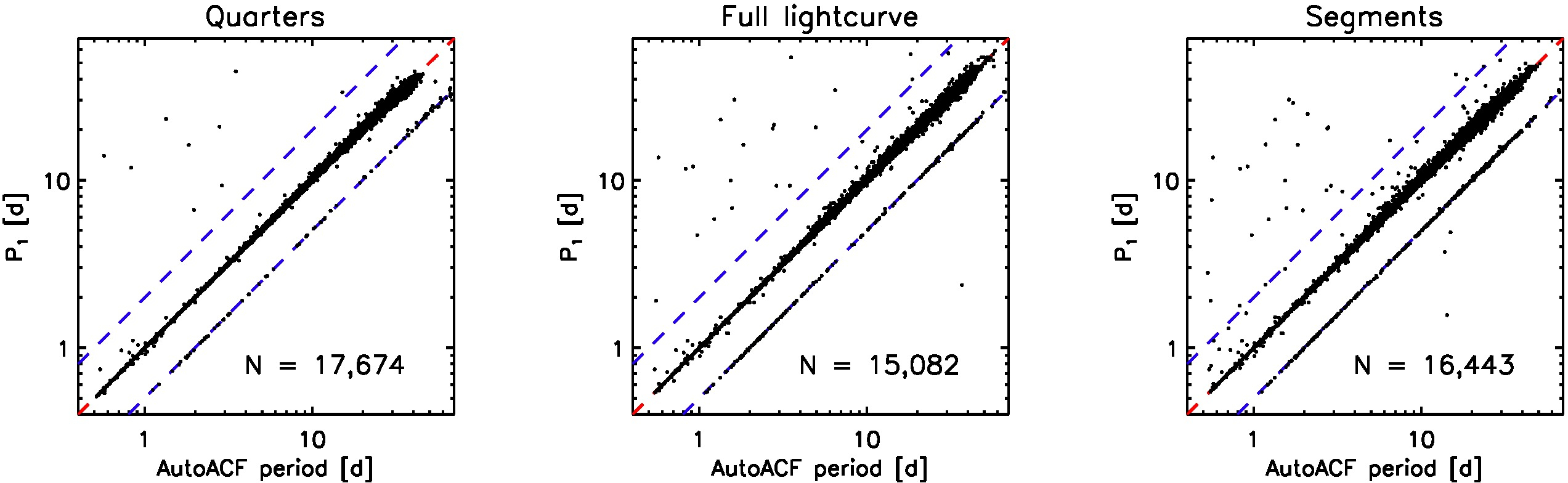}
  \caption{Comparison of rotation periods $P_1$ to the results of \citet{McQuillan2014}
  using different methods. The dashed red line denotes the 1:1 ratio, and the upper and
  lower dashed blue lines indicate the 2:1 and 1:2 period ratios, respectively. For
  each method the number of stars matching all criteria is given in the lower right
  corner of each panel.}
  \label{comp_periods}
\end{figure*}

% comparison of different methods to McQuillan (2014)
For the three different approaches (Sects.~\ref{quarters}--\ref{segments}) we compared our
measurements to the state-of-the-art rotation periods from \citet{McQuillan2014} in
Fig.~\ref{comp_periods}. In general, all methods show very good agreement. In each panel
the dashed red line shows the one-to-one period ratio, and the upper and lower dashed blue
lines indicate the 2:1 and 1:2 period ratios, respectively. The left panel shows average
periods from the quarters Q1--Q14, using the green, red, and blue dots from
Fig.~\ref{goodper}. We find \nPQ stars matching the two samples. Thereof, more
than \nPQperc\,\% of our measurements lie within 10\,\% of those from
\citet{McQuillan2014}. A small fraction of \nPQalias\,\% alias periods was found. The
middle panel shows rotation periods derived from the analysis of the full light curve.
\nPfull stars are matching, \nPfullperc\,\% thereof do not differ by more than 10\,\%, and
\nPfullalias\,\% alias periods were found. The right panel shows measurements from the
individual segments. \nPseg stars are matching, thereof \nPsegperc\,\% within 10\,\%, and
\nPsegalias\,\% alias periods were detected.

\begin{figure}
  \resizebox{\hsize}{!}
  {\includegraphics{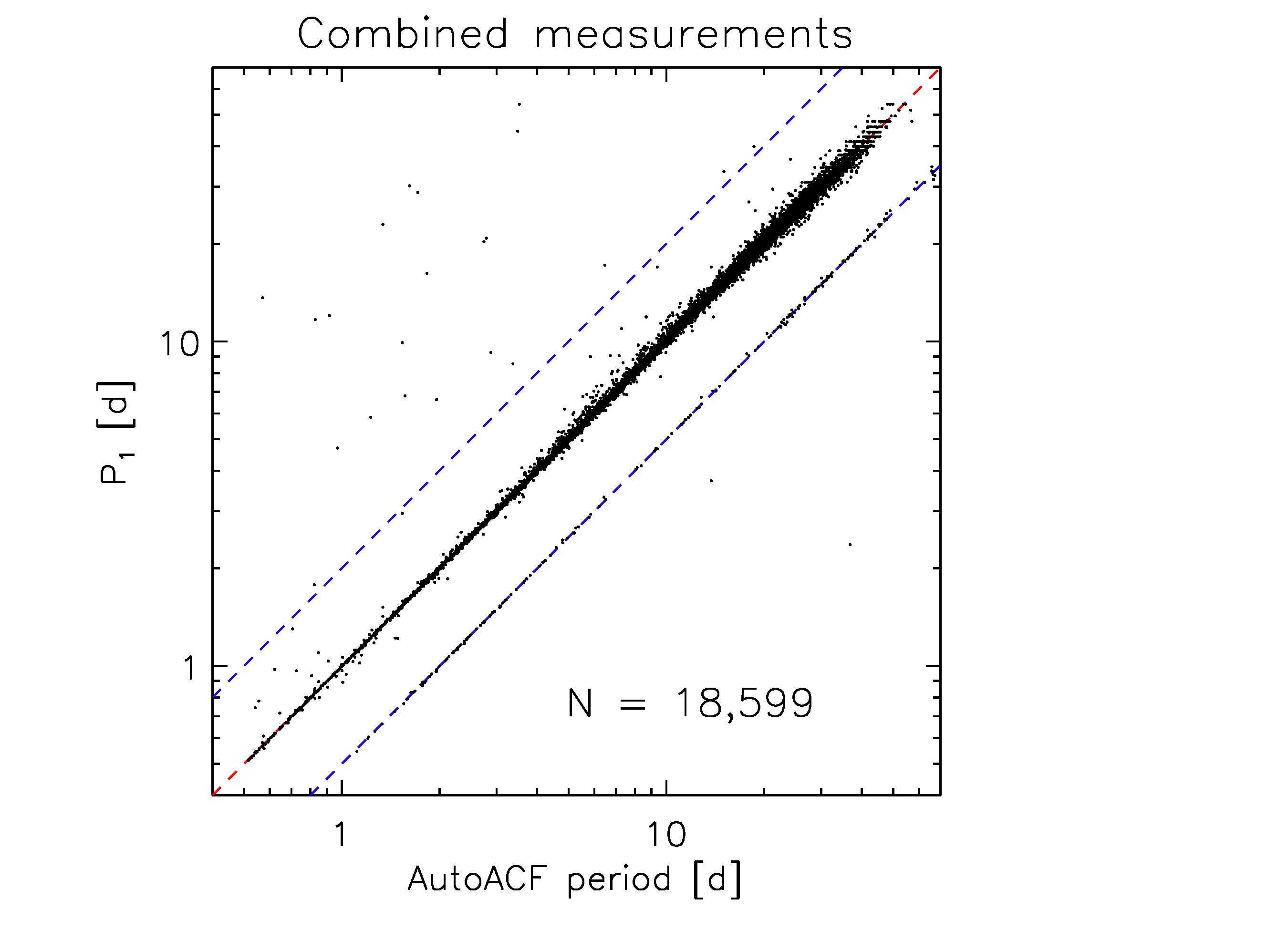}}
  \caption{Rotation period measurements combining the three different approaches from
  Fig.~\ref{comp_periods}, and comparing them to AutoACF periods from
  \citet{McQuillan2014}.}
  \label{combined}
\end{figure}

% combined measurements
Combined period measurements from the methods above are shown in Fig.~\ref{combined}. We
computed the median of the periods $P_1$ derived from each method, allowing for a median
absolute deviation (MAD) of one day for periods shorter than twenty days, and a MAD of two
days for longer periods. In total, we find \nPcomp stars matching with the sample of
\citet{McQuillan2014}. \nPcompperc\,\% of the periods lie within 10\,\% of each other. In
contrast to \citet{McQuillan2014} our sample contains stars hotter than 6500\,K, and also 
a few stars which do not belong to their periodic sample. Since we are searching for DR 
in the following section, we do not restrict our sample to stars matching with
\citet{McQuillan2014}, but consider all stars with measured period $P_1$ within the above
MAD limits, additionally satisfying $\log g > 3.5$ and $R_{\rm var}>0.3\,\%$.

\subsection{Differential rotation}\label{DR}
% general problems measuring DR
As shown in the previous section, rotation periods can be detected in a straight forward
way by picking the highest periodogram peak and applying certain selection criteria. The
situation is different when searching for differential rotation, acting as a perturbation
to the main rotation period. We usually interpret the detection of a second period in the
periodogram analysis as an indication of DR. Unfortunately, each method from
Sects.~\ref{quarters}--\ref{segments} yields different results. For a certain star it
occurs that one method returns a second period, whereas the other one does not. Even if
all three methods yield a second period for a certain star, the periods found can differ a
lot, depending on the different selection criteria and the frequency resolution. For that
reason we define mean values for the DR of a certain star in the following.

% definition of mean values Pmin & Pmax
We pick the minimum and maximum of the set of significant periods, individually for 
Sects.~\ref{full} and ~\ref{segments}. If both methods yield a minimum and maximum period
for a certain star, we compute the mean value, and call these periods $P_{\rm min}$ and
$P_{\rm max}$, respectively. We refrain from calculating minimum and maximum periods from
individual quarter measurements owing to their lower frequency resolution. Thus, we define
the relative and absolute horizontal shear $\alpha := (P_{\rm max} - P_{\rm min})/ P_{\rm
max}$ and d$\Omega := 2\pi\,(1/P_{\rm min} - 1/P_{\rm max})$, respectively, as a measure
of DR. In the following, we show how these two quantities correlate with rotation period
and effective temperature. We only consider stars exhibiting a variability range $R_{\rm
var}>0.3\,\%$, which was defined as a lower activity limit in \citet{Reinhold2013}.

\begin{figure}
  \resizebox{\hsize}{!}{\includegraphics{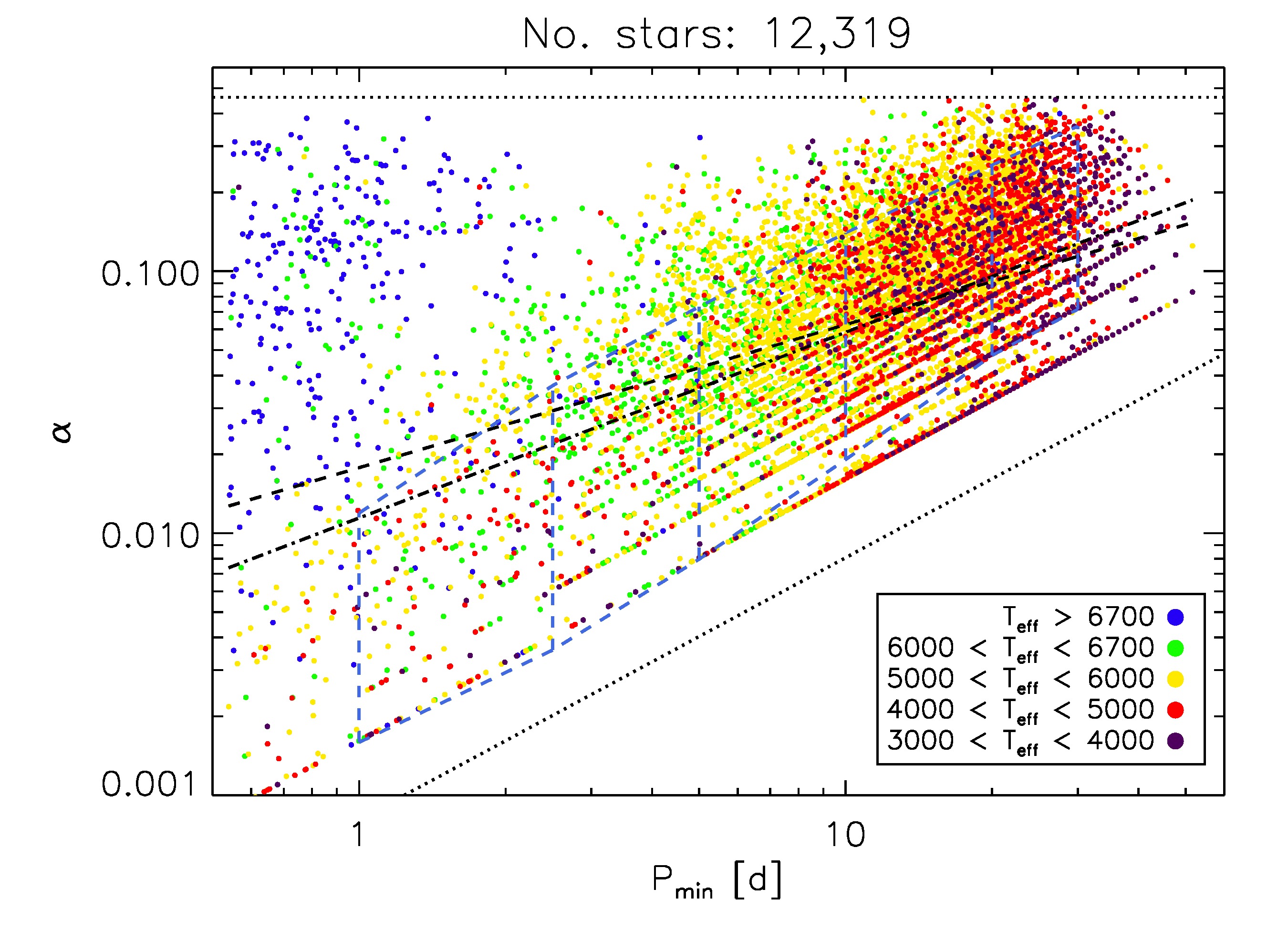}}
  \caption{Minimum period $P_{\rm min}$ versus relative differential rotation $\alpha$.
  The colors represent different temperature bins, the dotted black lines marks our
  detection limits. The dashed blue area shows theoretical predictions from
  \citet{Kueker2011} for models with $0.3 M_\odot$ (bottom) to $1.1 M_\odot$ (top). We
  find that $\alpha$ increases with rotation period for stars cooler than 6700\,K. The
  black lines show linear fits to our measurements, discarding stars hotter than 6700\,K
  (dash-dotted line) and using all data points (dashed line), respectively.}
  \label{Pmin_alpha}
\end{figure}

% Pmin vs. alpha
Fig.~\ref{Pmin_alpha} shows the relative shear $\alpha$ as a function of the minimum
period $P_{\rm min}$. For stars cooler than 6700\,K we find that $\alpha$ increases with
rotation period. A contrary behavior is found for hot stars ($T_{\rm eff}>6700$\,K)
populating the upper left corner ($\alpha > 0.02$ and $P_{\rm min}<2$\,d). These short
period stars spread a wide range of $\alpha$, and clearly do not follow the overall
trend. We suggest that multiple period measurements in these stars might actually be due
to pulsations or rapid spot evolution, and should not be interpreted as DR. Owing to
magnetic braking, cool stars exhibit longer periods, on average, thus populating the upper
right part of Fig.~\ref{Pmin_alpha}. In the lower left corner ($\alpha < 0.01$ and 
$P_{\rm min}<3$\,d) a mixture of all temperatures is found, indicating young fast
rotating stars. Again, we warn the reader that these small $\alpha$ values can also be
mis-classified as spot evolution. The dashed blue area shows theoretical predictions from
\citet{Kueker2011} for models with $0.3 M_\odot$ (bottom) to $1.1 M_\odot$ (top). These
simulations agree well with our observations of relative shear increasing with rotation
period. Furthermore, the hot stars are not covered by the model predictions, supporting 
our conclusion above. Almost 78\,\% of the fast rotators ($P_{\rm min}<2$\,d) exhibit
rotation periods very stable in time. These very stable periods are considered separately
in Sect.~\ref{stable}.

We compared our findings to previous measurements from \citet{Hall1991} who also found an
increase of the relative shear with rotation period. The dash-dotted and dashed black
lines show linear fits to our measurements in log-log space yielding a relation
$\alpha\propto P_{\rm min}^c$ with $c=0.71$ discarding the hot stars (blue data points)
and $c=0.55$ using all data points, respectively. The first value is consistent with the
result from \citet{Hall1991} who found $c=0.79\pm0.06$.

We also compared our results to measurements from \citet{Donahue1996}. These authors found
a relation between the mean rotation period $\langle P \rangle$ and the observed period
spread $\Delta P$ according to $\Delta P \propto \langle P \rangle^{1.3\pm0.1}$. Keeping
in mind that $\alpha\propto\Delta P/\langle P \rangle$, we find $\Delta P \propto \langle
P \rangle^{c+1} = \langle P \rangle^{1.71}$. This value is slightly bigger than the value
of $1.3$ found by \citet{Donahue1996}, regardless of whether discarding or keeping hot
stars. The discrepancy might be explained by the fact that these authors only considered
FGK stars, whereas we also incorporate M stars in our sample, usually exhibiting large
values of $\alpha$.

\begin{figure}
  \resizebox{\hsize}{!}{\includegraphics{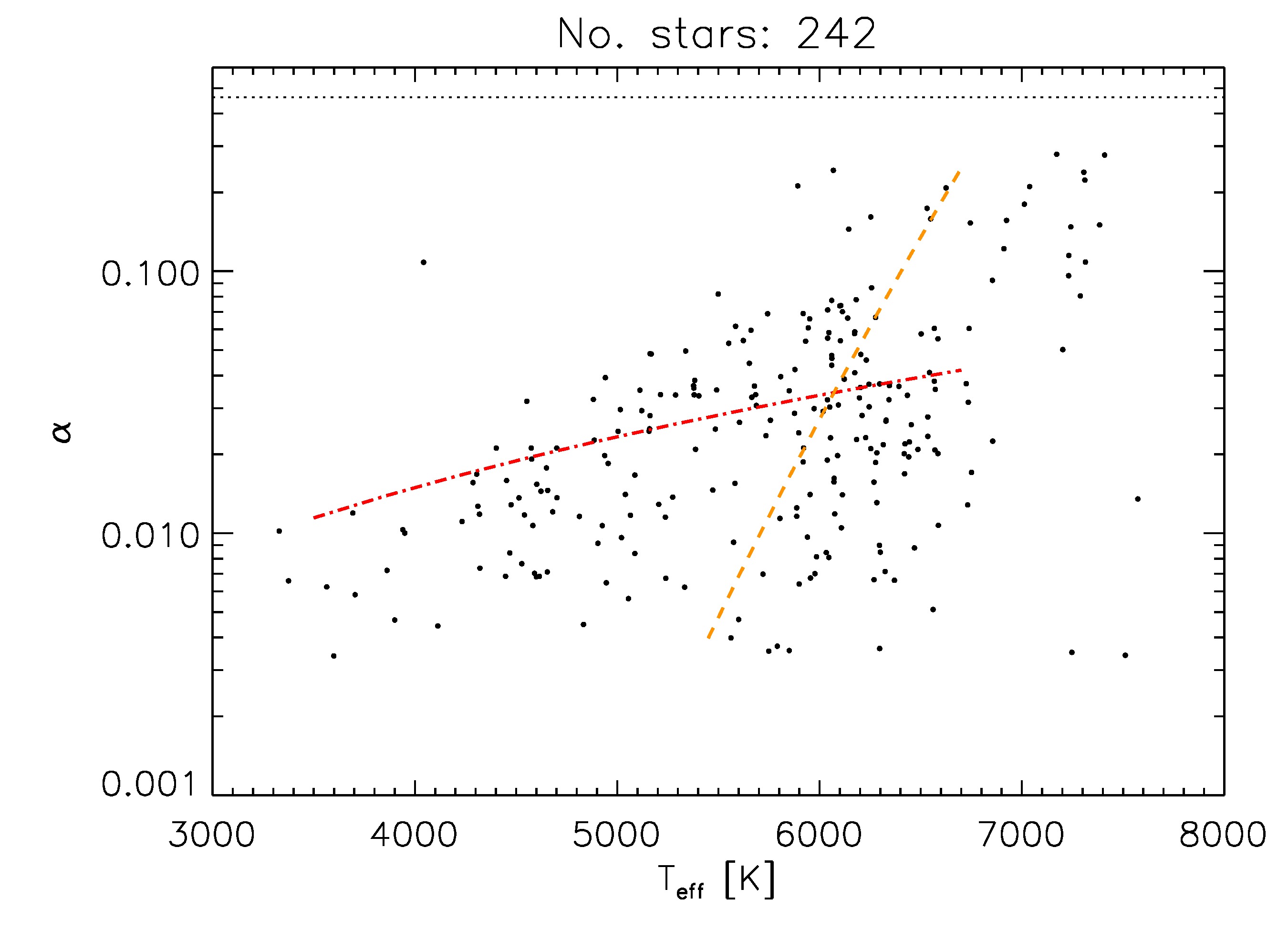}}
  \caption{$T_{\rm eff}$ versus $\alpha$ for stars satisfying $2\rm\,d<P_1<3\rm\,d$. The
  dash-dotted red and dashed orange curve show models from \citet{Kueker2011}. The
  dotted black line marks the upper limit of $\alpha$. No distinct correlation between
  $T_{\rm eff}$ and $\alpha$ was found.}
  \label{Teff_alpha}
\end{figure}

% Teff vs. alpha
Fig.~\ref{Teff_alpha} shows the dependence of the relative shear on effective temperature
for a subset of stars satisfying $2\rm\,d<P_1<3\rm\,d$. The dash-dotted red and dashed
orange curves show theoretical predictions from \citet{Kueker2011} using a model with an
equatorial rotation period of $2.5$ days. Between 3000--6000\,K the red curve matches the
observations quite well, although the slope of the curve is too small. Stars hotter than
6000\,K are well represented by the slope of the orange curve, although our observations
are offset towards higher temperatures. We are using revised temperatures from
\citet{Huber2014}, being on average 200\,K hotter for hot stars and 200\,K cooler for cool
stars, compared to previous temperatures from the KIC, which might explain this offset. We
do not want to stress the comparison of our observations to the models because we do not
find a clear trend when using all rotation periods. This is consistent with previous work
\citep{Reinhold2013}, where only a shallow trend of $\alpha$ towards cooler stars has been
found. We now interpret this shallow increase of $\alpha$ as a consequence of the increase
of rotation period towards cooler stars.

\begin{figure}
  \resizebox{\hsize}{!}{\includegraphics{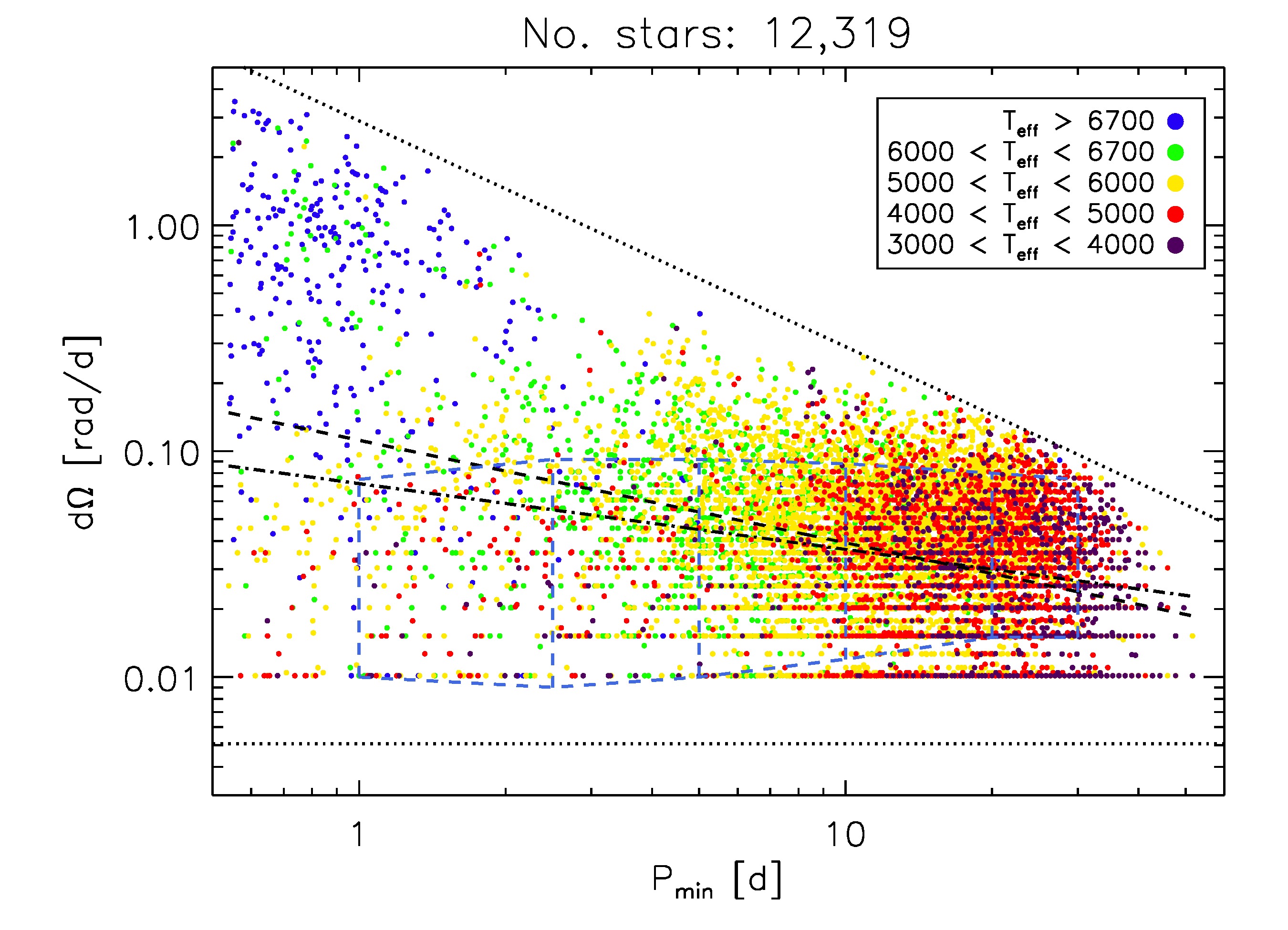}}
  \caption{Minimum period $P_{\rm min}$ versus absolute differential rotation d$\Omega$.
  The colors represent different temperature bins, the dotted black lines mark our
  detection limits. The dashed blue area shows theoretical predictions from
  \citet{Kueker2011}. We found that d$\Omega$ is almost independent of rotation period
  over a wide period range, but slightly increases towards short periods. Stars hotter
  than 6700\,K do not follow the overall trend. The black lines show linear fits to our
  measurements, discarding stars hotter than 6700\,K (dash-dotted line) and using all data
  points (dashed line), respectively.}
  \label{Pmin_dOmega}
\end{figure}

% Pmin vs. dOmega
In Fig.~\ref{Pmin_dOmega} we plot the absolute shear d$\Omega$ against rotation period
$P_{\rm min}$. We find that d$\Omega$ does not strongly depend on rotation period over a
wide period range. Towards fast rotators with periods on the order of a few days, the
absolute shear increases, although showing large scatter. Again, we find that the upper
left corner is populated by hot stars (s. Fig.~\ref{Pmin_alpha}), which are clearly
separated from the overall trend. Again, we warn the reader that these large d$\Omega$
values might not be associated with strong surface shear, but with different pulsation
frequencies or rapid spot evolution. The dashed blue area shows theoretical models from
Fig.~3 in \citet{Kueker2011} with $0.3 M_\odot$ (bottom) to $1.1 M_\odot$ (top), which
cover most of our measurements, but do not touch the hot stars with 
d$\Omega>0.1\,\rm rad\,d^{-1}$.

Equivalent to Fig.~\ref{Pmin_alpha} we applied a linear fit to our measurements yielding
d$\Omega\propto\langle P \rangle^c\propto\langle\Omega\rangle^{-c}$ with $c=-0.29$
discarding hot stars (dash-dotted line) and $c=-0.45$ using all data points (dashed line)
in Fig.~\ref{Pmin_dOmega}, respectively. We find good agreement with measurements from
\citet{Barnes2005} claiming d$\Omega\propto\langle\Omega\rangle^{0.15\pm0.10}$, confirming
the weak dependence of the absolute shear on the rotation rate.

\begin{figure*}
  \centering
  \includegraphics[width=17cm]{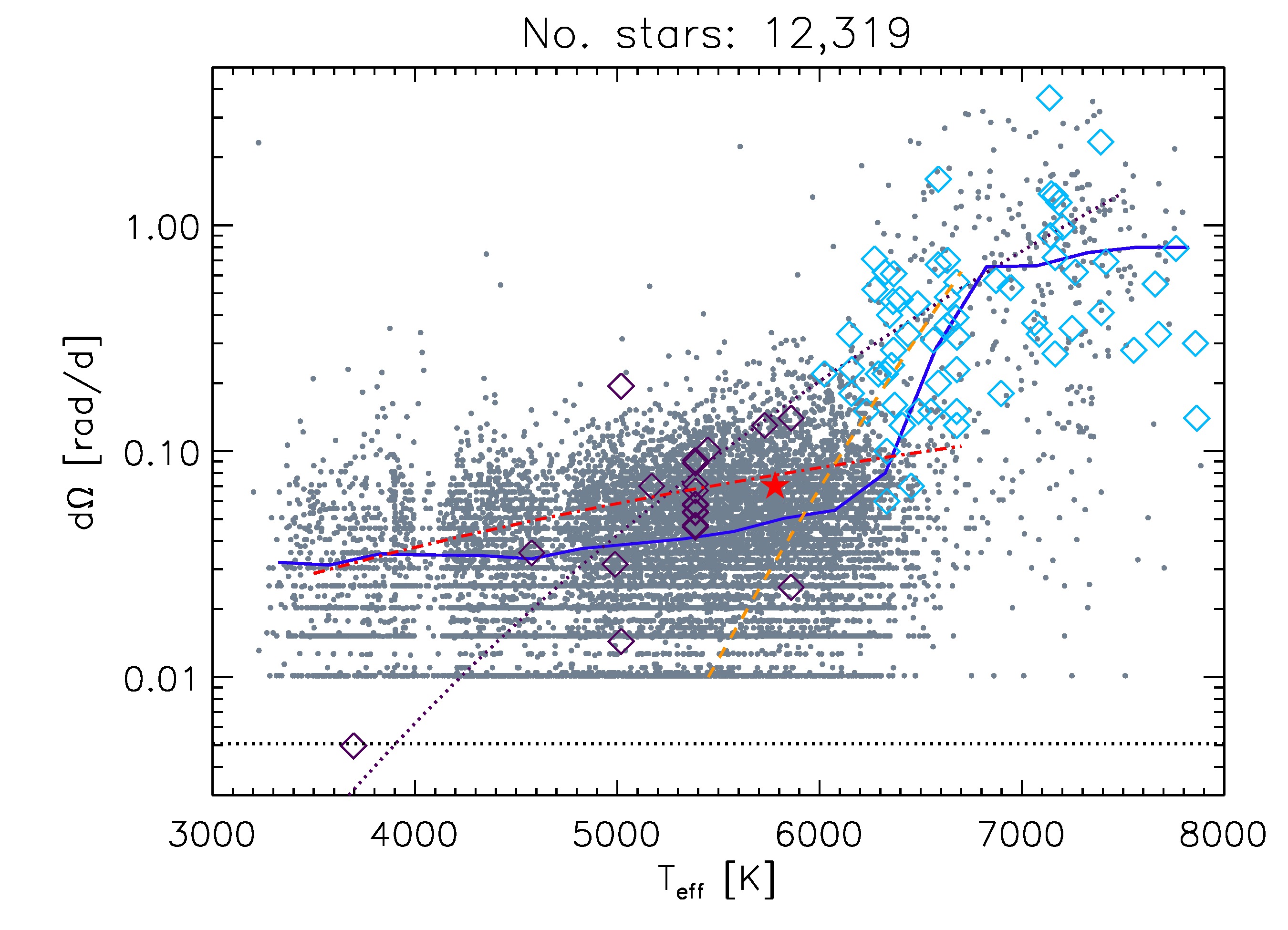}
  \caption{Effective temperature $T_{\rm eff}$ versus absolute shear d$\Omega$
  summarizing different measurements: Purple diamonds were taken from \citet{Barnes2005},
  the dotted purple curve from \citet{CollierCameron2007}. Light blue diamonds show
  measurements from \citet{Ammler2012}. Our measurements are shown as gray dots. The thick
  blue line represents the weighted mean of our measurements for 250\,K temperature bins.
  The dash-dotted red line and the dashed orange line show theoretical predictions from
  \citet{Kueker2011}. The dotted black line represents our detection limit, and the red
  star marks the position of the Sun. We find that d$\Omega$ shows weak dependence on
  effective temperature between 3300--6200\,K. Above 6200\,K d$\Omega$ strongly increases
  showing large scatter.}
  \label{Teff_dOmega}
\end{figure*}

% Teff vs. dOmega
The dependence of the absolute shear on effective temperature is shown in
Fig.~\ref{Teff_dOmega}. Our measurements (gray dots) suggest the existence of two distinct
regions with different behavior of d$\Omega$. From 3500--6000\,K the absolute shear
slightly increases showing weak dependence on temperature. Above 6000\,K d$\Omega$ steeply
increases with temperature, although showing large scatter. Data collected by
\citet{Barnes2005} is plotted as purple diamonds, and the corresponding fit from
\citet{CollierCameron2007} is shown as dotted purple line. These authors predicted a very
strong temperature dependence (d$\Omega\propto T_{\rm eff}^{8.9}$), which is not
supported by our measurements. Light blue diamonds show measurements\footnote{The
values were taken from Table~2 in \citet{Ammler2012} assuming an inclination of
$90^\circ$.} from \citet{Ammler2012} that are in good agreement with our findings.
Theoretical predictions from \citet{Kueker2011} are shown as dash-dotted red and dashed
orange curve. These authors found no sufficient match when fitting their results with one
polynomial over the whole temperature range. Therefore, they suggested a different
behavior of d$\Omega$ above and below $\sim$\,6000\,K. Their curves fit remarkably well
with the mean values of our measurements, shown as thick blue line. The matching gets even
better when using the old KIC temperatures, rather than the new values from
\citet{Huber2014}, which are 200\,K hotter (cooler) for the hot (cool) stars, on average.
The model from \citet{Kueker2011} is drawn from a star rotating with an equatorial period
of $P_{\rm eq}=2.5$\,d. Our observations contain various rotation periods, which might
explain the spread.
% The slight increase of d$\Omega$ with effective temperature between 3000--6000\,K is
% predicted by the lambda effect.
In Sect.~\ref{discussion} we discuss reasons for the observed spread, including cases with
multiple periods mis-interpreted as DR. The measured values of $P_{\rm min}$, $P_{\rm
max}$, $\alpha$, and d$\Omega$ are collected in Table~\ref{DR_table}. The measured period
differences are used as uncertainties for determining gyrochronology ages in the next
section.
% DR table
\begin{table}
  \centering
  \begin{tabular}{ccccccc}
\hline\hline
KIC & $T_{\rm eff}$ & $R_{\rm var}$ & $P_{\rm min}$ & $P_{\rm max}$ & $\alpha$ & d$\Omega$ \\
 & (K) & (\%) & (d) & (d) &  & rad/d \\
\hline
9771948 & 5678 & 1.63 & 15.115 & 15.689 & 0.037 & 0.015 \\
11122123 & 5960 & 1.17 & 9.883 & 10.955 & 0.098 & 0.062 \\
2558314 & 4157 & 0.82 & 24.790 & 28.171 & 0.120 & 0.030 \\
12317549 & 5793 & 1.06 & 7.377 & 7.996 & 0.077 & 0.066 \\
3103637 & 5332 & 0.71 & 26.372 & 33.500 & 0.213 & 0.051 \\
\hline
\end{tabular}

  \caption{DR measurements used in Figs.~\ref{Pmin_alpha}--\ref{Teff_dOmega}. The full
  table is available in machine-readable form in the online journal.}
  \label{DR_table}
\end{table}

\subsection{Stellar Ages}\label{stellar_ages}
\subsubsection{Gyrochronology}\label{gyro}
After measuring mean rotation periods $P$ and period variations $\Delta P$ for two-thirds
of the sample, we calculate stellar ages $t$ along with their uncertainties $\Delta t$
using different gyrochronology relations \citep{Barnes2007,Mamajek2008,Meibom2009},
hereafter abbreviated as B07, MH08, and MMS09, respectively. Ages are calculated according
to Eq.~3 in \citet{Barnes2007}:
% ages equation
\begin{equation}\label{age_eq}
  \log t = \frac{1}{n}[\log P - \log a - b\,\log X],
\end{equation}
with $X:=(B-V)_0-c$ and different fit parameters $a,b,c,n$ (see Table~\ref{gyro_table}).
Colors $(B-V)_0$ are obtained by converting $g-r$\footnote{Taken from the Kepler
Input Catalog (KIC).} to $(B-V)$ color using the relation from \citet{Jester2005} and
subtracting the excess $B-V$ reddening $E(B-V)$\footnotemark[8].
% ages table
\begin{table}
  \centering
  \begin{tabular}{ccccccc}
\hline\hline
Ref. & a & b & c & n & $(B-V)_0$ & P [d] \\
\hline
B07    & 0.7725 & 0.601 & 0.400 & 0.5189 & $\geq 0.42$ & $\geq 1.5$ \\
MH08   & 0.4070 & 0.325 & 0.495 & 0.5660 & $\geq 0.51$ & $\geq 1.5$ \\
MMS09  & 0.7700 & 0.553 & 0.472 & 0.5200 & $\geq 0.50$ & $\geq 1.5$ \\
\hline
\end{tabular}

  \caption{Fit parameters and period and color constraints used in the different
  gyrochronology relations.}
  \label{gyro_table}
\end{table}
Age uncertainties are calculated according to Eq.~10 in \citet{Barnes2007}:
\begin{equation}\label{age_uncert_eq}
  \Delta t=\frac{t}{n}\sqrt{(\Delta P/P)^2 + (b\,\Delta X/X)^2 + d\theta^2},
\end{equation}
% maybe use other g-r color uncertainties!
with $\Delta P:=(P_{\rm max}-P_{\rm min})/2$ being the uncertainty of the mean period $P$,
and $\Delta X$ being the $(B-V)$ difference between the conversion from \citet{Jester2005}
and \citet{Bilir2005}. Frequently, this difference is rather small, and no uncertainties
for the SDSS $g-r$ colors were found in the literature. Thus, we set a minimum uncertainty
of $\pm 0.01$\,mag to the $(B-V)$ colors with smaller uncertainties. The term $d\theta^2$
contains uncertainties of the fit parameters from Table~\ref{gyro_table} (compare Eq.~10
in \citealt{Barnes2007}). If DR was detected the period uncertainty dominates the age
uncertainty, especially for the slowly rotating stars with $\Delta P$ on the order of a
few days.

\begin{figure}
  \resizebox{\hsize}{!}{\includegraphics{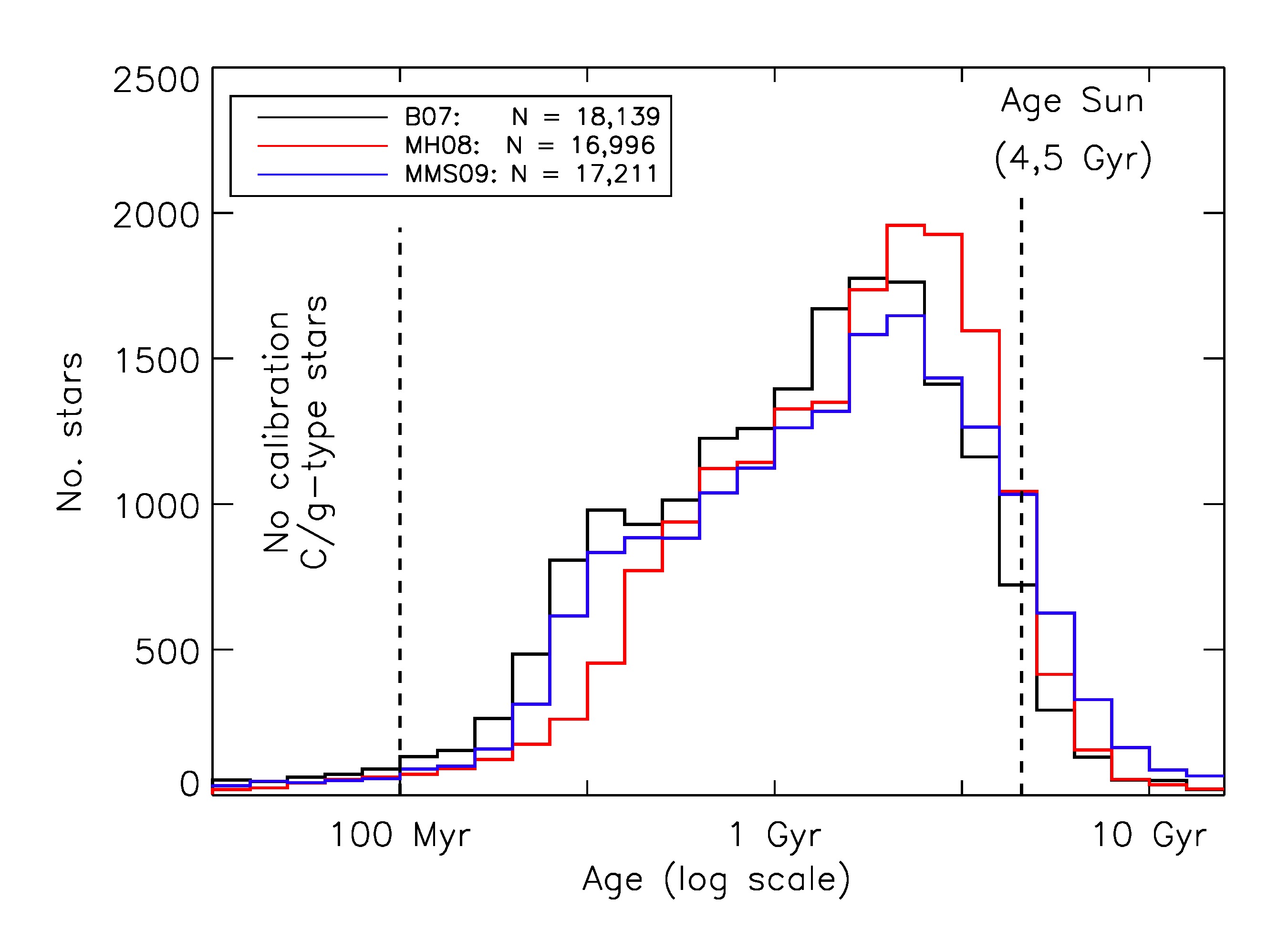}}
  \caption{Stellar age distributions derived from different gyrochronology relations:
  \citet{Barnes2007,Mamajek2008,Meibom2009}. Their calibration ranges are given in
  Table~\ref{gyro_table}. Stars with derived ages younger than 100\,Myr (left of the
  dashed black line) should be treated with caution.}
  \label{ages}
\end{figure}

% Age distributions
Distributions of the derived ages are shown in Fig.~\ref{ages}. For our stellar ages 
sample we retain stars that are not contained in the \citet{McQuillan2014} sample, but 
remove those with stable rotation periods (s. Sect.~\ref{stable}). The latter might not 
have spun down to the $I$-sequence yet (s. \citealt{Barnes2003} for terminology), or 
their rotation might be controlled by non-eclipsing companions. Additionally, we tighten 
our limit of the surface gravity to $\log g \geq 4.2$ to ensure that gyrochronology 
relations are only applied to dwarf stars.

The MH08 and MMS09 distributions contain less stars than the B07 distribution owing to 
the different color ranges (see Table~\ref{gyro_table}). The color and period constraints 
are needed to ensure that our field star sample obeys the same (or at least a similar) 
dependence of rotation period on color and age as the cluster stars used for calibration. 
The derived ages are collected in Table~\ref{age_table}, and their reliability is 
discussed in more detail in Sect.~\ref{discussion}.

% left and right edges
Derived ages younger than 100\,Myr should be treated with caution. These stars cover the
full $(B-V)_0$ range, but mostly exhibit rotation periods less than five days. They
likely have not yet converged to the $I$-sequence, and thus are not suitable for applying
gyrochronology relations. The right edge of the distribution with derived ages older than
10\,Gyr can also not be trusted since roughly half of the stars exhibit ages older than
the universe. These values result from a combination of long rotation periods
$(P > 20)$ days and effective temperatures $>5500$\,K, with either or both of them
being erroneous.

% numbers & percentages
Using the B07 distribution \ngoodage stars possess ages between 100\,Myr and 10\,Gyr.
Thereof, \ngoodageMattperc\,\% are younger than 4\,Gyr, in good agreement with
\citet{Matt2015} estimating $\sim$\,95\,\% comparing the sample from \citet{McQuillan2014}
to model predictions. Less than \badoldperc\,\% of the derived ages are greater than
10\,Gyr, and less than \badyoungperc\,\% of the B07 stars lie in the critical calibration
region younger than 100\,Myr, providing some confidence in the derived age distribution.
% ages histograms
% age Table
\begin{table*}
  \centering
  \begin{tabular}{cccccccccccccc}
\hline\hline
KIC & $R_{\rm var}$ & $T_{\rm eff}$ & $(B-V)_0$ & $\Delta (B-V)_0$ & $P$ & $\Delta P$ & $t_{\rm B07}$ & $\Delta t_{\rm B07}$ & $t_{\rm MH08}$ & $\Delta t_{\rm MH08}$ & $t_{\rm M09}$ & $\Delta t_{\rm M09}$ & flag \\
 & (\%) & (K) & (mag) & (mag) & (d) & (d) & (Myr) & (Myr) & (Myr) & (Myr) & (Myr) & (Myr) &  \\
\hline
7799260 & 2.14 & 5437 & 0.671 & 0.010 & 23.837 & 0.000 & - & - & - & - & - & - & ev \\
11135449 & 1.20 & 4731 & 1.032 & 0.010 & 31.783 & 0.775 & 2197 & 162 & 3154 & 364 & 2370 & 313 & - \\
3751292 & 4.21 & 6089 & 0.556 & 0.010 & 3.656 & 0.000 & 172 & 16 & 241 & 35 & 279 & 48 & - \\
8481446 & 0.39 & 5699 & 0.628 & 0.010 & 15.115 & 0.000 & 1710 & 134 & 1892 & 222 & 2213 & 259 & - \\
3757951 & 6.17 & 5345 & 0.677 & 0.010 & 1.918 & 0.006 & - & - & - & - & - & - & vs \\
\hline
\end{tabular}

  \caption{Rotation periods, $(B-V)_0$ colors, and gyrochronology ages using 
  different calibrations. If no ages are provided, either the period or the color lies 
  outside the valid calibration range (see Table~\ref{gyro_table}). \textit{Super-stable} 
  and \textit{very stable} stars (flagged as "ss" and "vs", respectively) were discarded 
  from the stellar ages sample. Furthermore, stars with $\log g < 4.2$ may have 
  \textit{evolved} off the main sequence (flagged as "ev"). For these stars we provide a  
  rotation period but no ages since gyrochronology relations are not calibrated for    
  evolved stars. Further table entries are the KIC number, the variability range 
  $R_{\rm var}$, and the effective temperature $T_{\rm eff}$.}
  \label{age_table}
\end{table*}

% activity-age relation
\subsubsection{Activity-age relation}
\begin{figure*}
  \centering
  \includegraphics[width=0.3\textwidth,height=0.2\textheight]{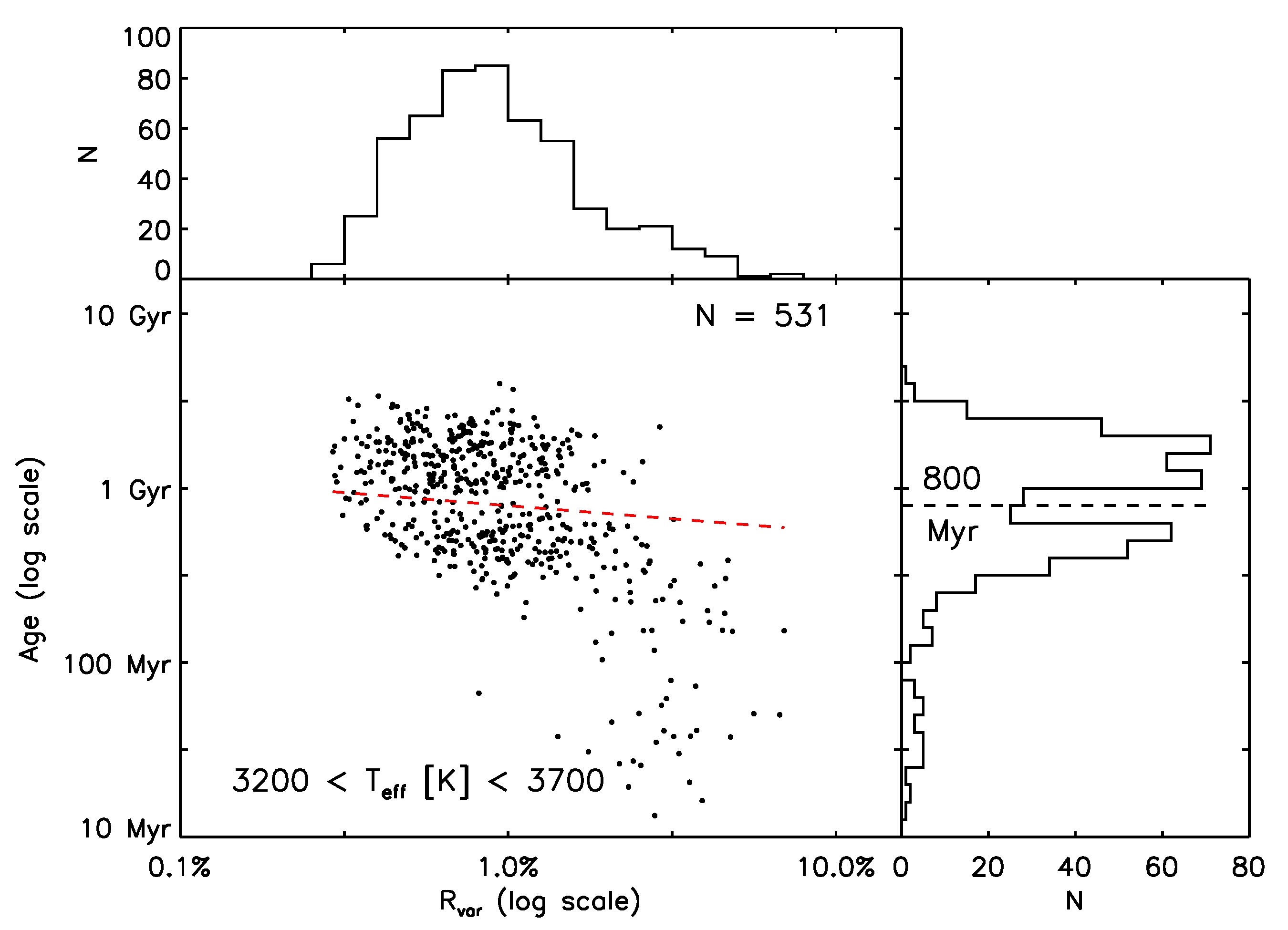}
  \includegraphics[width=0.3\textwidth,height=0.2\textheight]{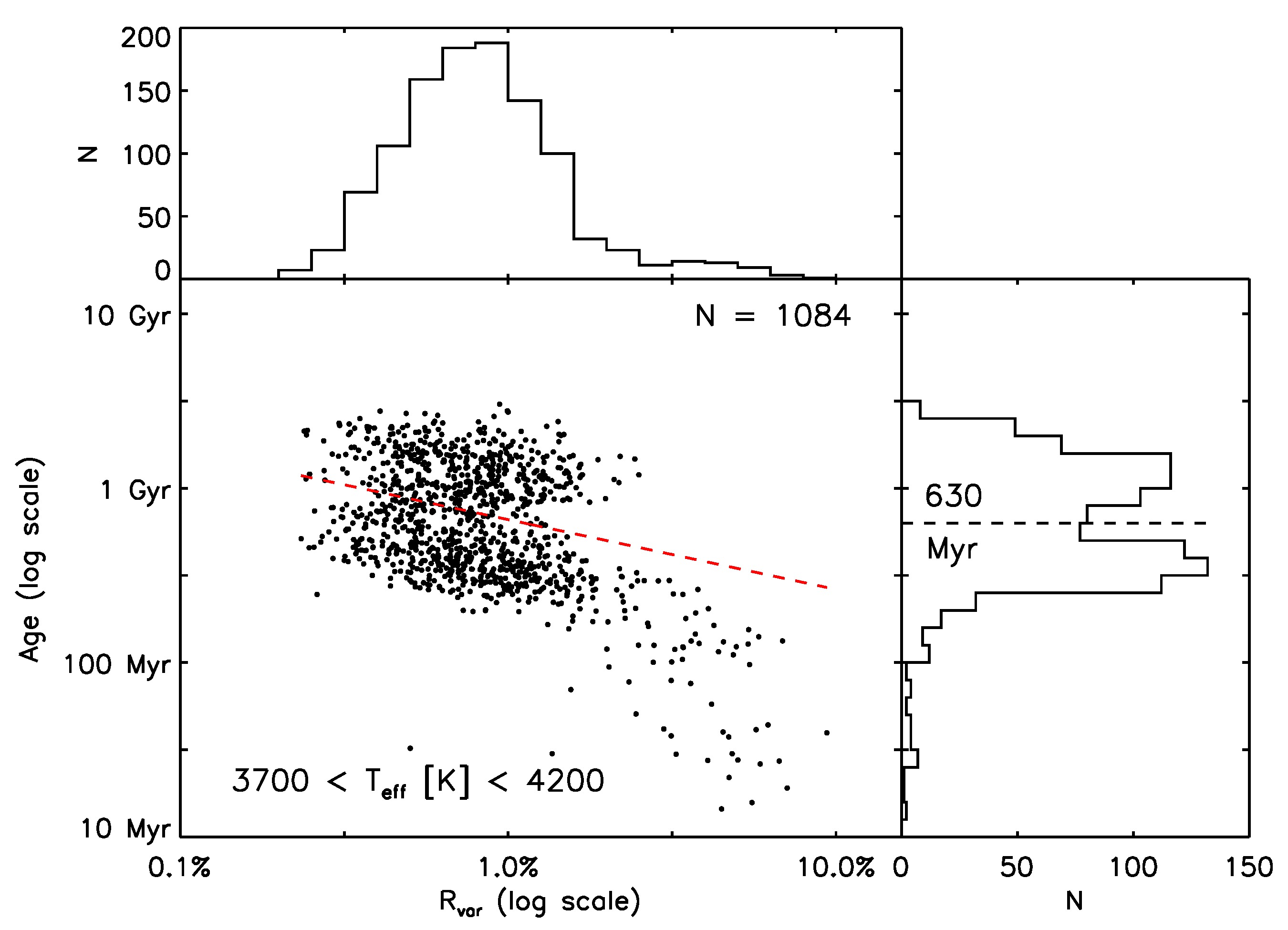}
  \includegraphics[width=0.3\textwidth,height=0.2\textheight]{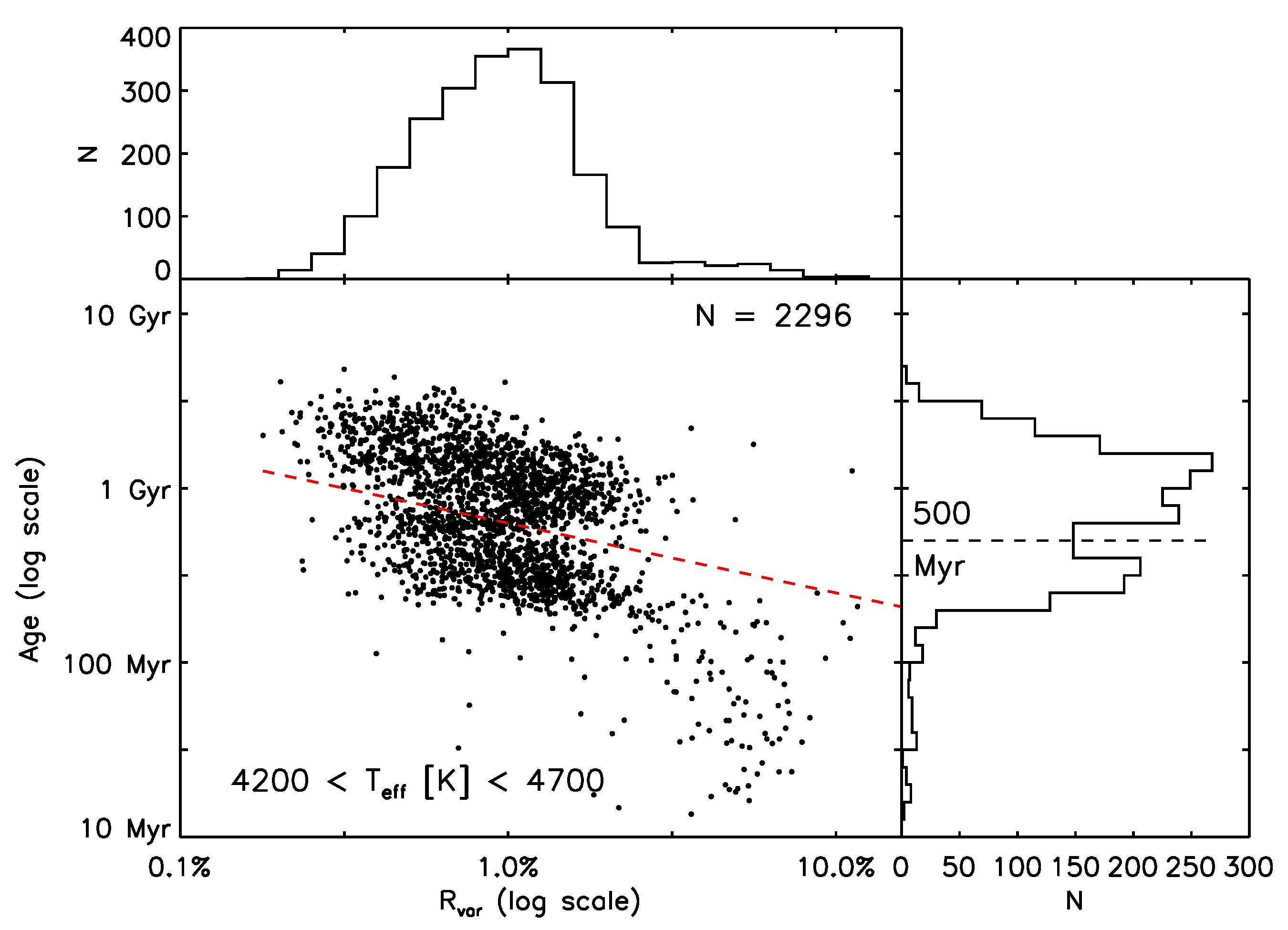}
  \includegraphics[width=0.3\textwidth,height=0.2\textheight]{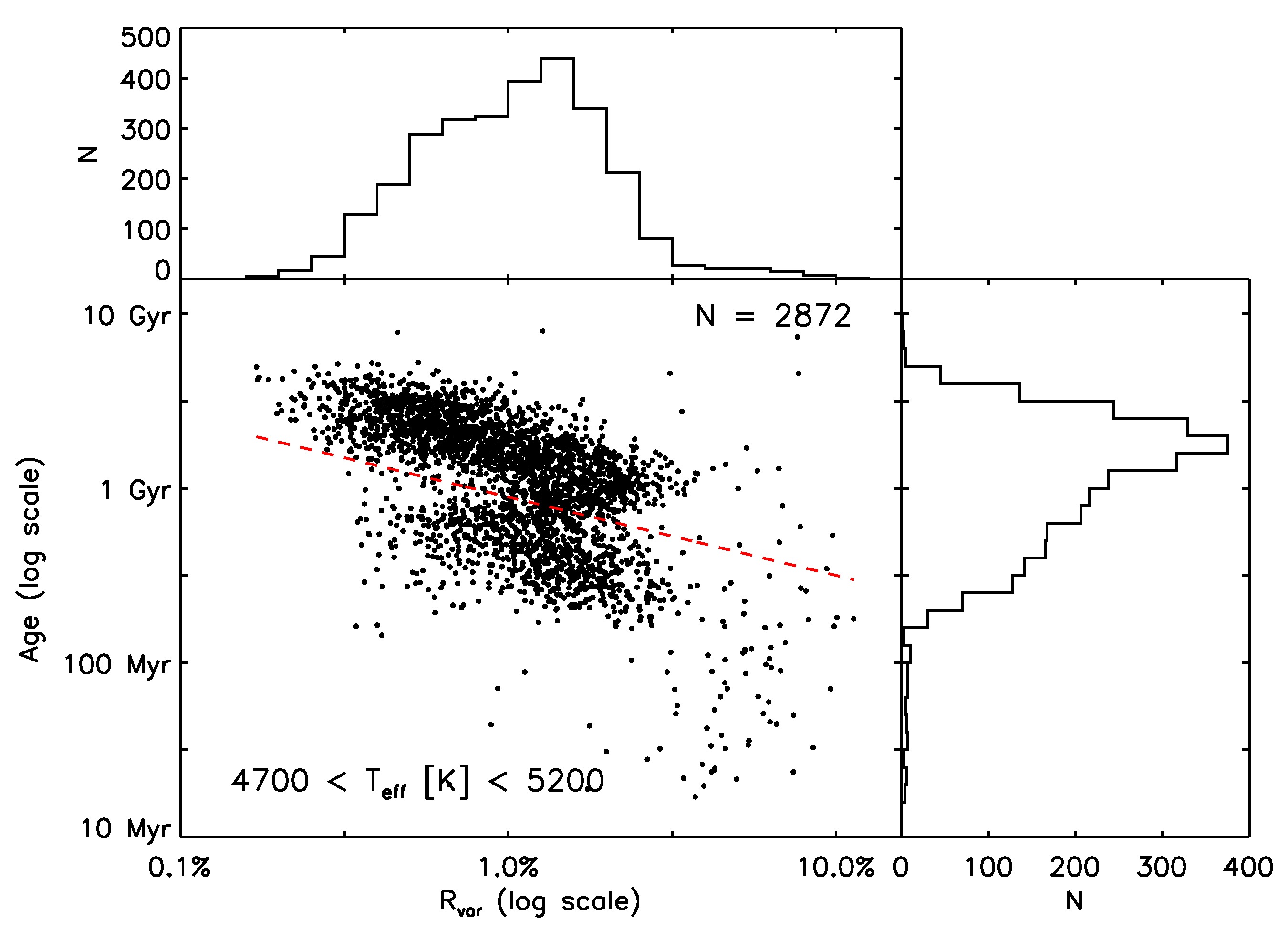}
  \includegraphics[width=0.3\textwidth,height=0.2\textheight]{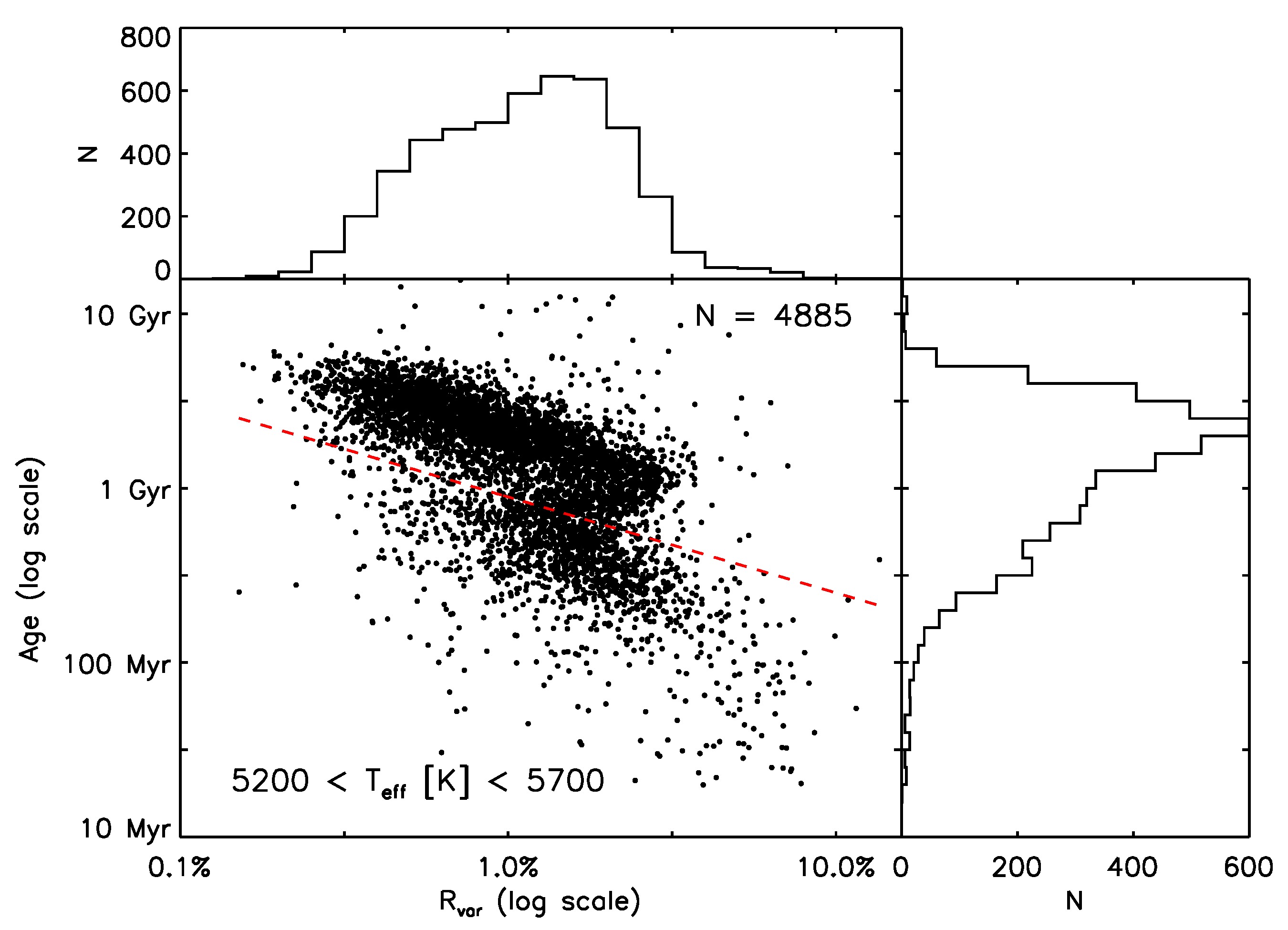}
  \includegraphics[width=0.3\textwidth,height=0.2\textheight]{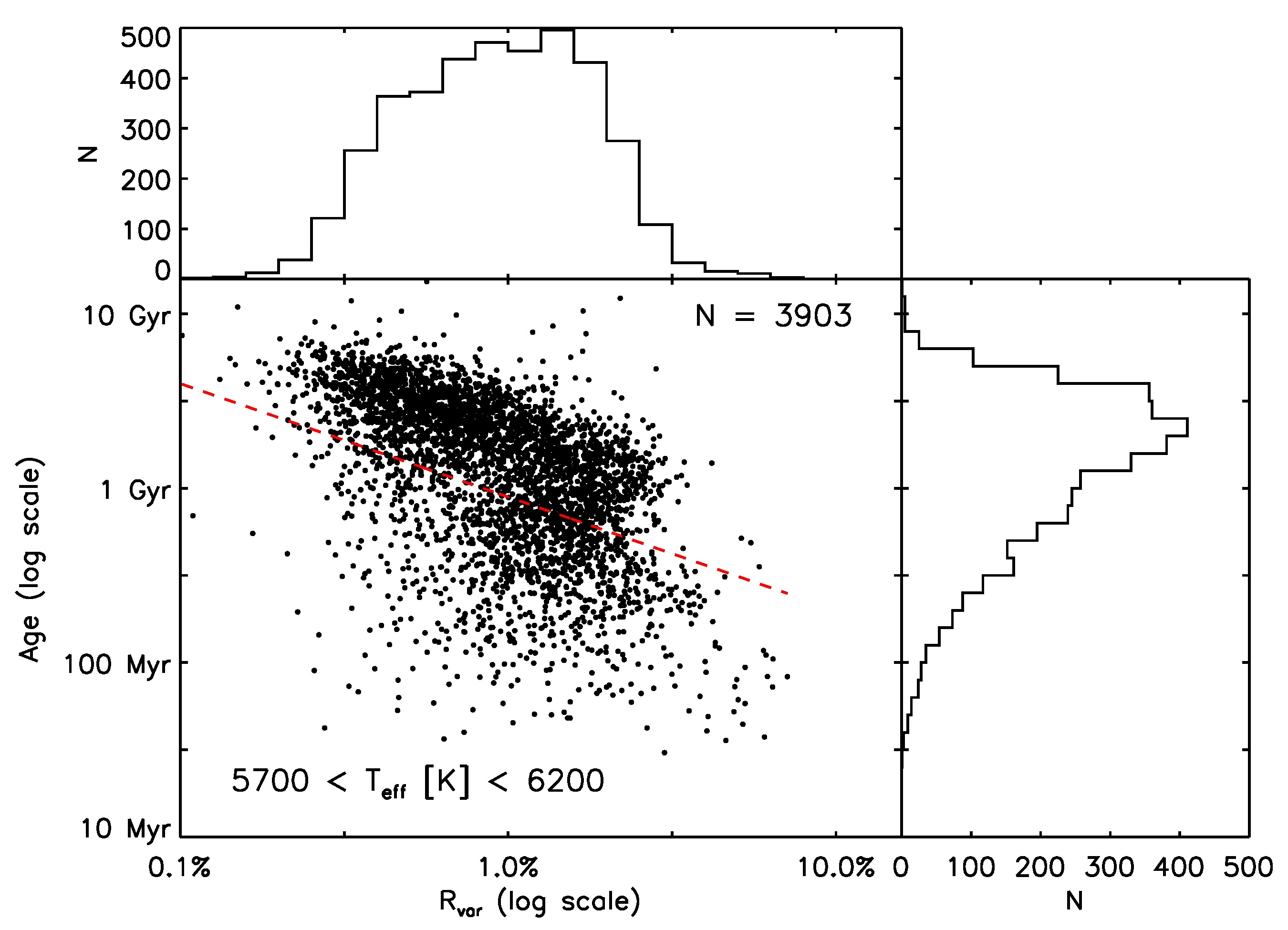}
  \caption{Gyrochronology ages plotted against the variability range for 500\,K
  temperature bins between 3200--6200\,K. The age distribution shows a bimodality between
  3200--4700\,K, which vanishes for hotter stars. The variability range $R_{\rm var}$
  decreases with age, becoming more emphasized towards hotter stars (increasing slope of
  dashed red line).}
  \label{range_age}
\end{figure*}

% motivation of activity-age relation
Inspired by Fig.~8 in \citet{Soderblom1991} we are interested in deriving a similar
activity-age relation. Unfortunately, spectra of Kepler stars are lacking, so we cannot
compare the derived ages to the established chromospheric activity measure $R'_{\rm HK}$.
Nevertheless, the variability range $R_{\rm var}$ can be used as an activity indicator in
a statistical sense. Using the B07 distribution we plot the age against the variability
range in Fig.~\ref{range_age}, which was inspired by Fig.~4 in \citet{McQuillan2014}.
The ages were derived only using periods lying closer than 10\,\% to periods found by
\citet{McQuillan2014}. From the upper left to the lower right, the temperature increases
from 3200--6200\,K in 500\,K intervals.
% age bimodality
The upper panels (3200--4700\,K) shows a bimodality of the age distribution, which
vanishes for hotter stars. This incident was first detected by \citet{McQuillan_Mdwarfs}
for the Kepler M dwarf periods, and confirmed later that the bimodality extends to hotter
stars \citep{Reinhold2013,McQuillan2014}. Moreover, the gap separating the two peaks
shifts towards younger stars with increasing temperature, starting at $\sim$\,800\,Myr for
the coolest stars (3200--3700\,K), and descends to 500\,Myr for stars between
4200--4700\,K.
% increasing trend; two samples
For each temperature bin the point distribution shows that the variability range
decreases with age, although each distribution exhibits large scatter in $R_{\rm var}$.
This general behavior is expected from the observation that young stars are, on average,
more active than old ones. In each panel two point clouds are visible. Thus, we separated
the two clouds into a so-called \textit{young} and \textit{old} sample. To guide the eye
we empirically drew a dashed red line between the two samples, with increasing slope
towards hotter temperatures, and varying offset in each panel. Stars lying below (above)
the dashed red line belong to the \textit{young} (\textit{old}) sample, respectively.
There are other ways of defining a \textit{young} and an \textit{old} sample, e.g., using
the horizontal dashed black line separating the two peaks of the bimodal distribution, but
we wanted to emphasize the correlation between activity and age. We were curious if the
parameters of these distinct samples substantially differ. Thus, we plotted histograms of
common stellar parameters such as $\log g$, FeH, brightness, and so on. Unfortunately, we
found no major difference in the parameters that could explain the evidence of the two
point clouds. However, we found a correlation between the ages shown and the corresponding
peak heights of the primary periodogram peaks. The peak heights of the \textit{young}
sample were, on average, higher than the peaks of the \textit{old} sample. We interpret
this result in terms of young stars being more active. Hence, their light curves are more
sinusoidal, because they suffer less by DR, spot evolution, or instrumental flaws, all
effects disrupting the light curve shape.
% activity-age plot

\subsection{Extremely stable periods}\label{stable}
% general remarks
From the analysis of the individual quarters we found periods $P_{\rm 1,Q}$ that are 
extremely stable in time. These stars have previously been shown in Fig.~\ref{goodper},
and are analyzed in more detail here. To quantify their temporal stability we computed the
median absolute deviation $\rm MAD(P_1):=\overline{|P_{\rm 1,Q} - \Pm|}$, and categorized
two groups of stable periodic stars: \textit{super-stable} stars with 
$\rm MAD(P_1)<0.001$\,d, and \textit{very stable} stars satisfying 
$\rm 0.001<MAD(P_1)<0.01$\,d. These stars are flagged in the last column of
Table~\ref{age_table}. We emphasize that these periods are stable over more than
three years of observation!

\begin{figure*}
  \sidecaption
  \includegraphics[width=12cm]{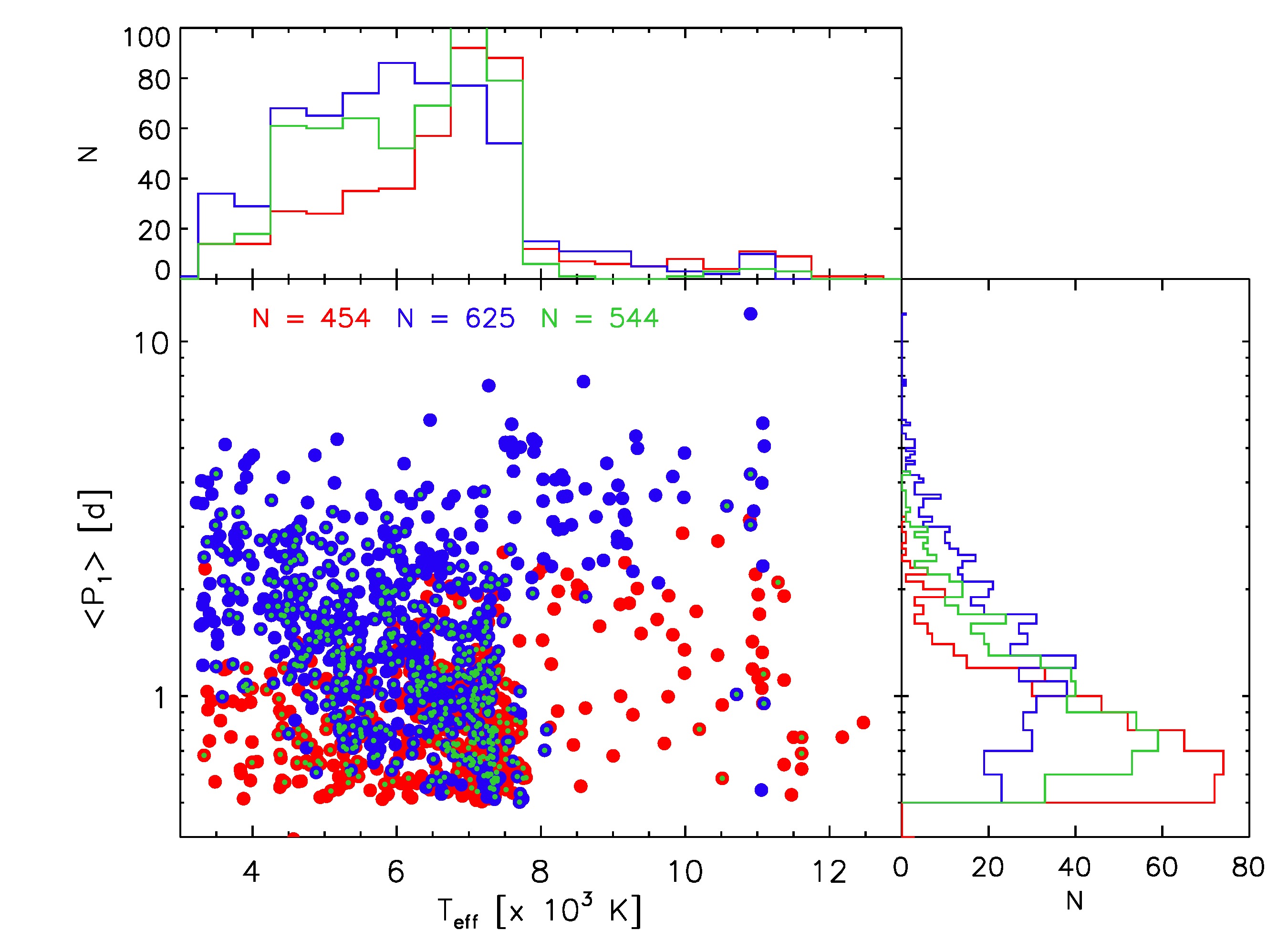}
  \caption{Stars with \textit{super-stable} periods ($\rm MAD(P_1)<0.001$\,d, red), and
  \textit{very stable} periods ($\rm 0.001<MAD(P_1)<0.01$\,d, blue). The inner green
  dots denote stars where a second period was found. The temperature and period
  distributions are shown in the top and right panel, respectively.}
  \label{stable_fig}
\end{figure*}

% description of the figure
Figure~\ref{stable_fig} shows the periods and effective temperatures of both groups.
\textit{Super-stable} and \textit{very stable} periods are shown as red and blue dots,
respectively. Inner green dots denote stars where a second period was found. The
temperature and period distributions are shown in the upper and right panel, respectively.
The period distribution shows that most of the \textit{super-stable} stars exhibit short
periods less than one day, with a mean period of $\Pm=0.95$\,d. The \textit{very stable}
stars extend to longer periods up to $\sim$\,12\,d, with a mean of $\Pm=1.90$\,d. Stars
with a second period were found in both groups, but only between 0.5--4.2 days. The
temperature distribution shows that the vast majority of both groups exhibit temperatures
less than 8000\,K, with a mean of $\sim$\,6800\,K for the \textit{super-stable} stars.
The \textit{very stable} stars are, on average, 700\,K cooler. Both distributions populate
the full temperature range, but stars with a second period were almost exclusively found
below 8000\,K. Between 8000--10,000\,K a dearth of second period stars was found, and only
few second period stars were found above 10,000\,K.

% possible explanation
There exist two general explanations for the observed stable periodic variability. One
possible explanation are stellar pulsations, which are known to be very stable in time.
The other process serving as an astronomical clock is synchronization by a companion. We
calculated the $\rm MAD(P_1)$ with the intention of disentangling these two
processes. In the period regime of 0.5--4 days $\delta$~Scuti, $\gamma$~Dor, and hybrid
pulsators thereof are expected. The boundaries of the so-called \textit{instability strip}
(i.e., the region in the Hertzsprung–Russell diagram populated by these pulsators)
is expected roughly between 6500--8800\,K. Surprisingly, most of the stable stars exhibit
temperatures less than 8000\,K. A fraction of them might actually be $\gamma$~Dor stars
between 6500--7000\,K, but there is no pulsation mechanism able to produce such stable
periods in the cool stars regime. Stable periods above 8000\,K are likely caused by
pulsations because spots are not necessarily expected for such hot stars.
% conclusion
Thus, we favor the conclusion that the periodicity of stars cooler than 6500\,K is
caused by spots on the stellar surface stabilized by non-eclipsing companions, either due
to interactions with another star or a close-in planet. Moreover, this hypothesis is
supported by the observation that stars with multiple periods (indicative for DR) were
mostly found below 8000\,K.
% all figures in one

\section{Discussion}\label{discussion}
\subsection{Rotation}
\begin{itemize}
\item We measured rotation periods for a statistically meaningful ensemble of stars. In
total, more than \nPcompround rotation periods we derived using different approaches, all
revealing very good agreement with the results from \citet{McQuillan2014}.
% maybe add Nielsen2013 comparison

\item As discussed in the previous section, we found \nstable periods extremely stable
in time, with a median absolute deviation less than 0.01\,d. For stars cooler than 6500\,K
binarity is the favored explanation. In our total sample (see Table~\ref{age_table}) we
have 5124 stars with $P_1<10$\,d and $T_{\rm eff}<6500$\,K. 573 of these stars exhibit
stable periods, which corresponds to a percentage of $\sim$\,11.2\,\%.
\citet{vanSaders2013} state that tidally synchronized binaries are fast rotators with
periods less than 10 days and that they contaminate the field with 4\,\%, compared to
11\,\% in the Hyades \citep{Duquennoy1991}. Our rate is almost three times higher than the
expected percentage of tidally locked field star binaries, but comparable to the expected
percentage in the Hyades. One possible explanation may be the existence of non-transiting
planets, which are able to decrease the angular momentum loss rate due to magnetized winds
\citep{Cohen2010}. Additionally, pulsating stars cooler than 6500\,K might contribute to
this rate. Other explanations for stable rotation might be different dynamos.
\citet{Brown2014} suggests the so-called \textit{Metastable Dynamo}, where stars are born
rapidly rotating with weak coupling to the wind. Other dynamo mechanisms might generate
strong magnetic fields leading to long spot lifetimes.
\end{itemize}

\subsection{Differential rotation}
\begin{itemize}
% spread of DR values
\item Exact values of $\alpha$ and d$\Omega$ are hard to determine, and depend on the
particular threshold used. Depending on which periods are selected as $P_{\rm min}$ and
$P_{\rm max}$, the absolute values of $\alpha$ and d$\Omega$ can differ a lot.
Nevertheless, each method from Sects.~\ref{quarters}--\ref{segments} produces the same
correlations for $\alpha$ and d$\Omega$ with temperature and period. Combined measurements
from the different approaches, as described at the beginning of Sect.~\ref{DR}, were used
to provide average values of $P_{\rm min}$ and $P_{\rm max}$. As discussed by the example
of Fig.~\ref{Teff_dOmega} the observed spread in d$\Omega$ can only partially be explained
by theory. \citet{CollierCameron2002} and \citet{Donati2003} found a large spread in
d$\Omega$ ranging from $0.046-0.091\,\rm rad\,d^{-1}$ for the active star AB~Dor.
Evolving spot configurations might be an explanation. Hence, all measurements here should
be interpreted in a statistical way.

\item Although the derived values are method dependent, the statistical averages are not.
Our results are in good agreement with previous measurements \citep{Hall1991,Donahue1996}
and theoretical predictions \citep{Kueker2011}.

\item Simulations of spotted stars exhibiting rapid spot evolution are able to generate
beat-shaped light curves, multiple periodogram peaks, and therefore able to mimic DR. This
interpretation was not considered so far. We think that spot evolution may play a role,
but we have no way to discriminate between these two phenomena.
\end{itemize}

\subsection{Stellar Ages}
\begin{itemize}
% calibration of gyrochronology relations
\item All gyrochronology relations have been calibrated by ground-based observations of
open clusters and Mount Wilson stars. The stars used in the different calibrations exhibit
different period and color ranges. Age calibration was performed using the Sun as an age
anchor. Thus, the relations are not tested for stars older than the Sun. Applying these
relations to stars with a wider period and color range might lead to less accurate ages. 

% \item Moreover, the photometric accuracy and duty cycle of ground-based observations is
% not sufficient to resolve different rotation rates. The periods used for calibration
% should rather be considered as mean rotation periods of the stars. Thus, we also use 
mean
% rotation periods in the relations, even when $P_{\rm min}$ and $P_{\rm max}$ are
% available. However, these periods supply a good estimate of the period uncertainty, 
which
% in the end dominates the age uncertainty.

% reliability of ages
\item Gyrochronology ages are most reliable between 500--2500\,Myr. Depending on their
braking efficiency some stars younger than 500\,Myr may not have converged to the
$I$-sequence yet. Thus, their periods may not be suitable for the use of gyrochronology.
The calibration becomes even worse for stars with derived ages less than 100\,Myr. Such
stars might belong to the $C$-sequence, obeying a physically different behavior. We do not
trust derived ages younger than 100\,Myr or older than 10\,Gyr. 

% subgiant contamination
\item Subgiants and main sequence stars obey a different rotation-age relationship
\citep{Garcia2014}. In this study we attempt to exclude evolved stars by setting a lower 
limit to the surface gravity of $\log g \geq 4.2$. Unfortunately, Kepler does not provide 
stellar luminosity classification, so our sample might be contaminated by subgiants. 
Contamination might be as large as 35\,\% for field stars as pointed out by 
\citet{vanSaders2013}.

% two point clouds
\item The existence of the two point clouds in Fig.~\ref{range_age} for stars with 
$3200 < T_{\rm eff} < 4700$\,K is a matter of debate. \citet{McQuillan2014} suggested 
that the bimodality of the age distribution can be understood in terms of two distinct 
star formation events in the solar neighborhood. Stellar ages are correlated with the 
variability range in the sense that young stars are more active, on average. 
Interestingly, this trend becomes more distinct towards hotter stars, as indicated by the 
dashed red line in Fig.~\ref{range_age}, which lacks an explanation so far.

% asteroseismology ages
\item A comparison of gyrochronology and asteroseismology ages is challenging. Much
progress was recently made (see, e.g., \citealt{Angus2015}). Most rotation period and
asteroseismology samples do not overlap because strong activity, essential for achieving
spot rotation periods, damps mode excitation \citep{Chaplin2011}. Furthermore, Kepler
lacks bright stars, which are needed for asteroseismology. Upcoming missions will
hopefully change this unpleasant situation in the near future.
\end{itemize}

\section{Summary}\label{summary}
We re-analyzed the sample of \oldsample stars from \citet{Reinhold2013} using Q1--Q14
data. Good agreement was found with previous rotation periods and measurements from
\citet{McQuillan2014}. We searched for deviations from the mean rotation period using the
Lomb-Scargle periodogram in different approaches, aiming to detect multiple significant
periods, and assigning them to surface differential rotation. The general trends of the DR
with period and temperature, observed in \citet{Reinhold2013}, could all be confirmed
although individual measurements of $\alpha$ and d$\Omega$ may differ due to the different
frequency resolution of the full time series and the 90-days time base of a single
quarter. In general, the new measurements are in very good agreement with previous
observations \citep{Hall1991,Donahue1996} and theoretical predictions \citep{Kueker2011}.
Stellar ages were derived from gyrochronology relations provided by different authors,
with uncertainties that are dominated by the period spread. A bimodal age distribution was
found between 3200--4700\,K, vanishing for hotter star. The derived ages show a
correlation with the variability range serving as an activity indicator. Furthermore, we
found \nstable stars exhibiting a very stable period, with a median absolute deviation
less than $0.01$\,d. The almost constant periods of the hot stars may be explained by
pulsations, whereas the stability of the cooler star ($T_{\rm eff}<6500$\,K) may be
explained with synchronization of the orbital period of a non-eclipsing companion.

\acknowledgements
We would like to thank the referee for providing very constructive suggestions, which led 
to significant improvements to the paper, especially to the differential rotation 
discussion. We acknowledge support from Deutsche Forschungsgemeinschaft Collaborative 
Research Center SFB-963 ``Astrophysical Flow Instabilities and Turbulence'' (Project A18).

\bibpunct{(}{)}{;}{a}{}{,} % to follow the A&A style
\bibliography{biblothek}
\bibliographystyle{aa}

\end{document}